\documentclass[fleqn,usenatbib]{mnras}
\usepackage{lmodern}
\usepackage{fix-cm}
\usepackage{bm}
\usepackage{newtxmath}
\usepackage{natbib}
\usepackage[utf8]{inputenc}
\usepackage[T1]{fontenc}
\usepackage{graphicx}
\usepackage{amsmath}
\usepackage{mathtools}
\usepackage{array}
\usepackage{multirow}
\usepackage{calc}
\usepackage{url}

\usepackage{amssymb}
\usepackage{siunitx}
\usepackage{xspace}
\usepackage{enumitem,calc}
\usepackage{placeins}
\newlength{\symbollabel}
\newlist{symbollist}{description}{1}
\setlist[symbollist]{%
  font=\normalfont,
  labelsep=0.5em,
  labelwidth=\symbollabel,
  leftmargin=\dimexpr\labelwidth+\labelsep\relax, 
  itemsep=0.2ex,
  topsep=0.3\baselineskip,
  align=parleft 
}
\setlist[itemize]{leftmargin=*,align=left}
\setlist[enumerate]{leftmargin=*,align=left}
\usepackage{microtype}
\usepackage{tabularx}
\usepackage{booktabs}
\usepackage{relsize}
\usepackage{textgreek}
\usepackage{hyperref}
\usepackage{newtxmath} 
\usepackage{bm}
\DeclareMathSizes{5.5}{5.5}{5.5}{5.5}
\hypersetup{colorlinks=true, linkcolor=blue, citecolor=blue, urlcolor=blue}

\usepackage{enumitem}
\newlength{\symwidth}
\setlist[symbollist]{%
  font=\normalfont\itshape,         
  labelsep=0.5em,                   
  leftmargin=\dimexpr\symwidth+0.5em\relax,
  itemsep=0.2ex plus 0.2ex,         
  parsep=0pt, topsep=0.4\baselineskip
}
\graphicspath{{./}{./Figures/}}

\usepackage{newtxtext,newtxmath}

\usepackage{capt-of}

\usepackage{titlesec}

\allowdisplaybreaks           
\let\std\relax
\newcommand{\std}{\mathop{\mathit{std}}\nolimits}

\makeatletter
\renewcommand*{\@fnsymbol}[1]{\arabic{#1}} 
\renewcommand\thefootnote{\arabic{footnote}} 
\renewcommand\@makefnmark{%
  \hbox{\@textsuperscript{\footnotesize\thefootnote}}
}
\makeatother
\makeatletter
\renewcommand\@makefntext[1]{%
  \noindent\hbox to 1.5em{\hss\textsuperscript{\footnotesize\thefootnote}}#1}
\makeatother

\title[\emph{MESS}: Multi-Epoch Spectroscopic Solver]{\emph{MESS}: Multi-Epoch Spectroscopic Solver for Detecting Double-Lined Systems}

\author[G. Nachmani et al.]{
Gil Nachmani,\thanks{E-mail: \href{mailto:gilnachmani@mail.tau.ac.il}{gilnachmani@mail.tau.ac.il}}
Simchon Faigler 
 and Tsevi Mazeh
\\
School of Physics and Astronomy, Faculty of Exact Sciences, Tel Aviv University, Tel Aviv 69978, Israel\\
}

\date{Accepted XXX. Received YYY; in original form ZZZ}
\pubyear{2025}

\begin{document}
\label{firstpage}
\maketitle

\begin{abstract}
We present \emph{MESS}, a fully automated algorithm for identifying and characterizing double-lined spectroscopic binaries ($\mathcal{SB}2$) in large databases of multi-epoch spectra. \emph{MESS} extends the two-dimensional TODCOR approach to a global multi-epoch formalism, deriving the radial velocities (RVs) of both components at each epoch while optimizing the templates jointly across all observations. Template optimization searches a continuous synthetic-spectra manifold spanning an eight-dimensional parameter space: effective temperature, surface gravity, and rotational broadening for each star, together with a common metallicity and the flux ratio. Single-lined spectroscopic binaries ($\mathcal{SB}1$) and single stars ($\mathcal{S}1$) are handled within the same framework by fitting one optimized template, with either epoch-dependent RVs ($\mathcal{SB}1$) or a single shared RV ($\mathcal{S}1$). Model selection among $\mathcal{S}1/\mathcal{SB}1/\mathcal{SB}2$ uses the Bayesian information criterion with an effective sample size that accounts for intra-spectrum correlations, and is complemented by the Wilson relation between the two RVs to infer the mass ratio and systemic velocity without a full orbital solution. We validate \emph{MESS} on 1500 simulated LAMOST MRS systems (SNR$=50$), with primary RV semi-amplitudes predominantly below the instrumental resolution, achieving an overall classification accuracy of $\sim95\%$. We also derive full orbital solutions for two $\mathcal{SB}2$ systems detected in our LAMOST analysis, including a faint-secondary case with flux ratio $\sim0.1$, and present example outputs for one $\mathcal{SB}1$ and three constant-velocity stars. A companion paper will report the survey-wide application to LAMOST DR11 and the resulting $\mathcal{SB}1/\mathcal{SB}2$ catalogs.
\end{abstract}


\section{Introduction}
\label{sec:intro}

Binary and multiple stars are fundamental to stellar astrophysics. Classic volume-limited surveys established that about half of the nearby solar-type primaries reside in multiple systems \citep{DuquennoyMayor1991,Raghavan2010}. At higher primary masses, binarity is even more prevalent \citep{Sana2012}. 
%
Radial-velocity (RV) surveys probe binaries with periods from days to years and capture their characteristics. Spectroscopic binaries (SBs) are central to stellar astrophysics, providing mass ratios and systemic velocities and, when combined with astrometry or photometry, model-independent masses and radii. However, efficient detection and characterization are hindered by line blending, rotational broadening, template mismatch, and heterogeneous data quality.

A key to the analysis of SBs is the 1D cross-correlation function \citep[1D-CCF;][]{tonrydavis1979} and its mask-based implementations \citep{baranne1996,pepe2000}, which underpin much of the precise RV literature.
This technique is well suited for single-lined spectroscopic binaries ($\mathcal{SB}1$s), systems for which the secondary spectrum is too faint to be observed in the spectrum. However, for double-lined spectroscopic binaries ($\mathcal{SB}2$s), systems in which the lines of both components can be detected, the 1D-CCF is not ideal. Direct-modeling techniques can be very helpful \citep[e.g.,][]{elbadry2018b,kovalev22, kovalev23,kovalevChenHan22, kovalevChen24}. 
Complementary techniques include broadening functions \citep{rucinski1992}, least-squares deconvolution \citep[LSD;][]{donati1997}, full spectral disentangling in Fourier/wavelength space \citep{hadrava1995,simonsturm1994,ilijic2004}, and maximum-likelihood/template-matching RVs \citep{anglada2012}. 

Another approach is \emph{TODCOR}, a two-dimensional correlation function that is better suited for deriving the two RVs of each spectrum.
The 1D-CCF
mostly tracks the bright component of the spectrum and therefore can bias the primary correlation peak and often misses the faint one,  
especially at low flux ratios. 
TODCOR
explicitly models the observed spectrum as a combination of \emph{two} templates and
solves for both velocities simultaneously. The 2D joint maximization of
the correlation therefore de-blends the observed spectrum in velocity space. Even when the lines overlap, the 2D
surface exhibits an unambiguous peak at the two velocities
whereas a 1D-CCF is flattened or
skewed.

Beyond de-blending, the 2D surface encodes the local curvature and orientation of the correlation peak,
which informs on the covariances between the two velocities and the velocity uncertainties. 
Altogether, TODCOR improves
detection and measurement of faint companions while providing more reliable error geometry than
is available from any single 1D correlation. 


Large multi-epoch surveys (e.g. LAMOST (Large Sky Area Multi-Object Fiber Spectroscopic Telescope): \citealt{zhao2012}; APOGEE: \citealt{majewski2017}; Gaia RVS: \citealt{katz2023}; Gaia-ESO: \citealt{merle2020}; 4MOST: \citealt{dejong2019}) demand automated, statistically principled pipelines that \emph{jointly} decide among possible models for the observed data, return per-epoch velocities with uncertainties, and propagate model selection to orbital fits. In Gaia DR3 specifically, the Non-Single Star (NSS) content includes $>$\num{181000} spectroscopic-binary solutions \citep{Arenou2023}.
%
The broader DR3 NSS context comprises $\sim 8\times10^{5}$ multiplicity solutions
spanning astrometric, spectroscopic, and eclipsing binaries \citep[e.g.][]{Arenou2023}.
These large surveys motivate $\mathcal{SB}2$-oriented tooling. 

Here we introduce \emph{MESS}---Multi-Epoch Spectroscopic Solver, which extends TODCOR with \emph{continuous} template optimization and explicit multi-epoch model selection.
Its goal is to infer robust per-epoch radial velocities by exploiting all available spectra simultaneously and to select the best-fitting model among: single stars ($\mathcal{S}1$), single-lined binaries ($\mathcal{SB}1$), and double-lined binaries ($\mathcal{SB}2$). This is achieved through:
\begin{itemize}
\item a multi-epoch correlation metric that combines per-epoch 1D 
(for $\mathcal{S}1/\mathcal{SB}1$)
or 2D TODCOR surfaces (for $\mathcal{SB}2$) 
correlation functions
into a single score;
\item a continuous search over synthetic spectra, interpolated in $(T_{\rm eff},\log g,Z,V\sin{i})$, to optimize the underlying templates for all epochs jointly;

\item model selection via the Bayesian information criterion (BIC), supplemented by simple, survey-specific override rules based on RV amplitudes, Wilson-plot diagnostics, and phase-coverage tests to refine the final $\mathcal{S}1/\mathcal{SB}1/\mathcal{SB}2$ classification.
\end{itemize}

We first present in section~\ref{sec:todcor} the TODCOR method for a single spectrum, and demonstrate its capabilities when applied to one LAMOST red-arm exposure. 
Section~\ref{sec:mess} presents the philosophy and method of \emph{MESS}, including the model set for $\mathcal{S}1/\mathcal{SB}1/\mathcal{SB}2$ cases, continuous template optimization, multi-epoch scoring with an effective sample size, and Wilson-plot inference with rule-based overrides. 
Section~\ref{sec:validation} presents a simulation that verifies the performance of \emph{MESS}, and
Section~\ref{sec:examples}
demonstrates the potential of \emph{MESS} by analyzing a few LAMOST systems. 
Section~\ref{sec:disc} discusses sensitivity, limitations, and survey implications. The Appendices provide implementation details and derivations together with our simulation-based verification protocol.

\section{One-Spectrum TODCOR}
\label{sec:todcor}

RV inference from stellar spectra is based here on cross-correlation: when a single component dominates, one uses the canonical 1D cross-correlation function (1D-CCF) to measure the template’s Doppler shift; for double-lined spectra, a two-template generalization (TODCOR) models both components simultaneously, including their flux ratio, and yields a 2D surface over the two RV lags.

We work on a common log-$\lambda$ pixel grid. The observed spectrum $f$ and all templates $g$
are continuum-normalized and mean-subtracted. Following \citet{tonrydavis1979}, the 1D cross-correlation at lag $s$ is
%
\begin{equation}
C(f,g;s) \;\equiv\; \frac{\sum_{n} f_n\, g_{n+s}}{\sigma(f)\,\sigma(g)}\, ,
\label{eq:ccf1d}
\end{equation}
%
where $\sigma(\cdot)$ is the standard deviation across pixels on the working grid.
We will use Eq.~\eqref{eq:ccf1d} when assuming only one visible component in the spectrum
(i.e.,\ $\mathcal{S}1$ and $\mathcal{SB}1$ models).

The TODCOR algorithm \citep{zucker1994} handles two visible components by correlating  the spectrum against a
two-template model with flux ratio $\alpha$. With primary and secondary templates $g_1,g_2$, it
returns a 2D surface over Doppler shifts $(s_1,s_2)$:
%
\begin{equation}
R_\alpha(s_1,s_2)
\;=\;
\frac{C_1(s_1)+\alpha\,C_2(s_2)}
{\sqrt{\,1+2\alpha\,C_{12}(s_2-s_1)+\alpha^2\,}} \,,
\label{app:eq:Ralpha_recall}
\end{equation}
%
where $C_1(s_1)\equiv C(f,g_1;s_1)$, $C_2(s_2)\equiv C(f,g_2;s_2)$\\
and 
$C_{12}(s_2-s_1)\equiv C(g_1,g_2;s_2-s_1)$, 
as in Eq.~\eqref{eq:ccf1d}. 
We will use Eq.~\eqref{app:eq:Ralpha_recall} when assuming that two components are present in the observed spectrum.
The maximum of $R_\alpha$ at $(\widehat{s}_1,\widehat{s}_2)$ yields the RV estimates
$(\rm{v}_1,\rm{v}_2)$.

\subsection{Example: 1D-CCF vs.~the full 2D surface for one spectrum}
\label{sec:todcor:example}

To demonstrate the advantage of TODCOR, we present here an analysis of one co-added LAMOST 
red-arm
spectrum of LAMOST ID J114511.45+341926.0\footnote{Gaia ID 4031090178986430080} (hereafter J1145),
selected from the 
LAMOST DR10 Medium-Resolution Survey (MRS) multi-epoch catalog.\footnote{\url{https://www.lamost.org/dr10/}}
J1145 was observed 35 times and the full analysis of the system is presented below, in section \ref{sec:examples}.
 It turned out that J1145 has a low flux ratio between the secondary and primary, $\alpha \simeq 0.11$, so the secondary component contributes only weakly to the observed spectra.

\begin{figure}
  \centering
  \includegraphics[width=\columnwidth]{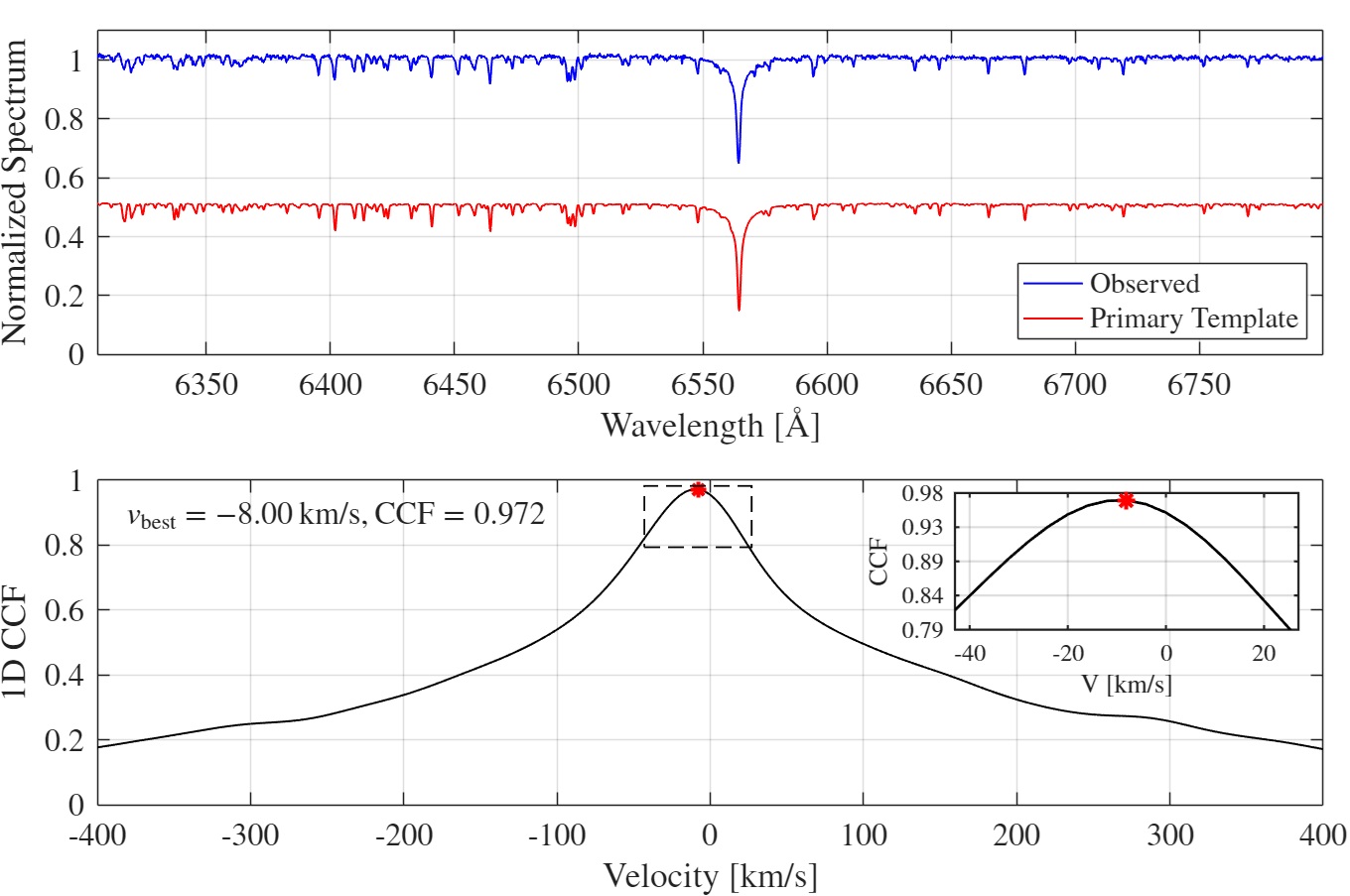}
  \caption{Single-epoch spectrum and 1D cross-correlation for J1145.
    \emph{Upper:} normalized observed spectrum (blue) at a representative epoch and the best-fitting
    \emph{primary} template (red; vertically offset by $-0.5$ in normalized flux for clarity) obtained from a 1D grid search over
    $(T_{\rm eff},\log g,Z,V\sin{i})$.
    \emph{Lower:} the corresponding 1D cross-correlation function (CCF) used to select both the
    template and the Doppler shift, showing a maximum at
    $v=-8\pm6 \,\mathrm{km\,s^{-1}}$.}
  \label{fig:todcor1D}
\end{figure}

Here we confine our analysis to the first spectrum obtained, using the templates found (see below) for this specific spectrum. 
A standard 1D cross-correlation function (CCF) against a single template fits the primary well, but entirely misses the secondary signal. 
Figure~\ref{fig:todcor1D} shows the 1D-CCF of one observed spectrum versus the best-fitting template. No hint of a secondary can be seen.
 
\begin{figure}
  \centering
  \includegraphics[width=1.0105\columnwidth]{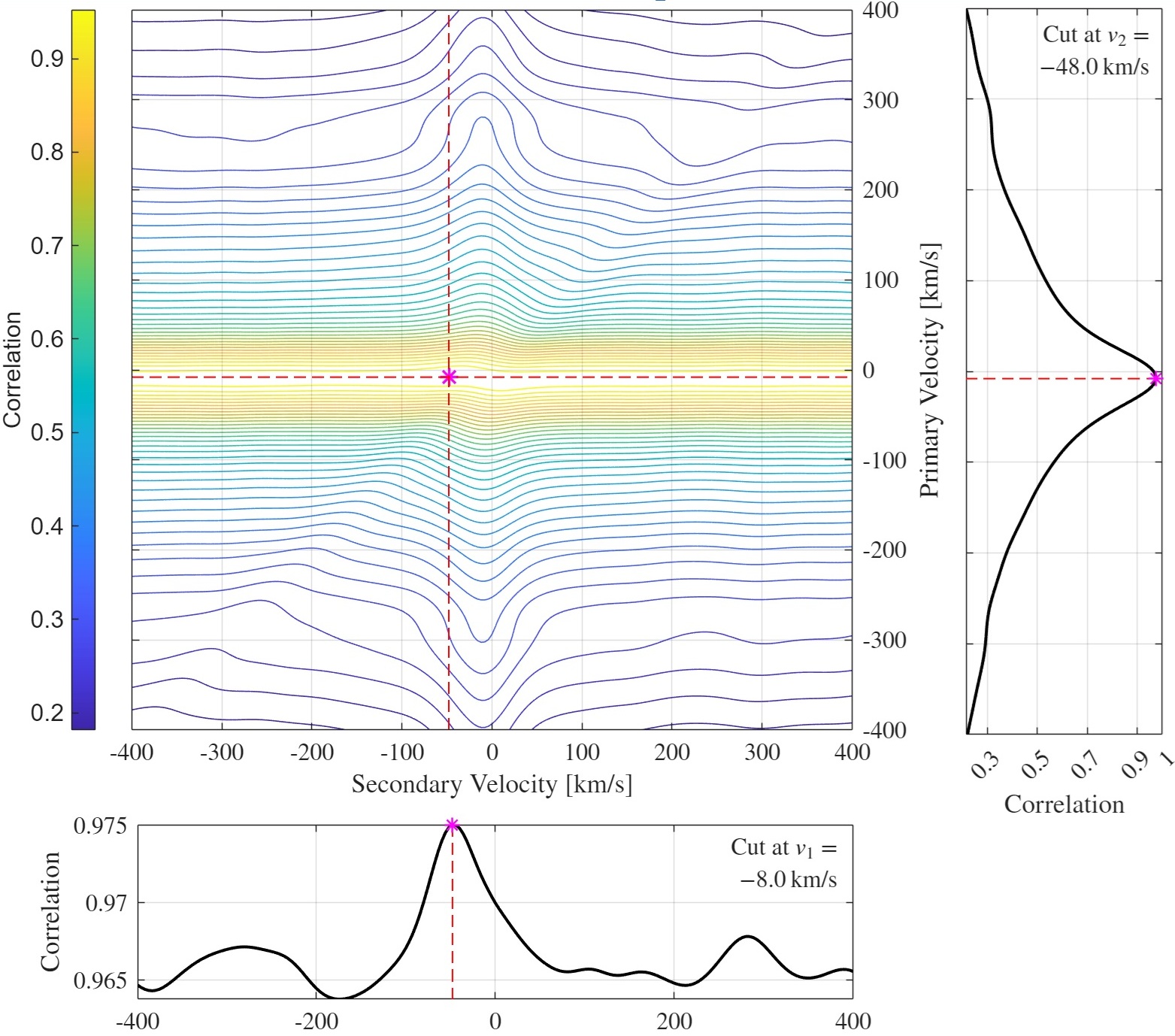}
  \caption{Two-dimensional TODCOR surface and orthogonal cuts for J1145 (same target as in figure~\ref{fig:todcor1D}).
    Left: the 2D correlation surface $R_\alpha(\rm{v}_1,\rm{v}_2)$; the magenta star marks the global maximum and red dashed lines indicate the orthogonal 1D sections.
    Right/bottom: 1D cuts through the peak along the $\rm{v}_1$ and $\rm{v}_2$ axes, respectively, illustrating the ridge orientation and blend-driven covariance.
    Best-fit orbital parameters and derived quantities are listed in table~\ref{tab:J1145_orbit}.}
  \label{fig:todcor2D}
\end{figure}

Figure~\ref{fig:todcor2D} presents the corresponding \emph{complete} 2D correlation map of the same observed spectrum, with constant-level contours and 1D cuts rendered alongside. The selected template parameters for this system appear in table \ref{tab:J1145_orbit}. The secondary cut (holding $\rm{v}_1=-8.0\, \mathrm{km\,s^{-1}}$ fixed and scanning over $\rm{v}_2$) shows that varying $\rm{v}_2$ perturbs only the faint component, and consequently, the correlation varies only slightly (from $0.975$ down to $0.965$). Nevertheless, the peak at $\rm{v}_2=-48 \, \mathrm{km\,s^{-1}}$ is clear and robust, while the \emph{primary} cut shows a high-contrast peak, rising from $\sim 0.3$ up to $0.975$, indicating how much the bright component dominates the spectrum. 
 In this specific case, even when the RV difference is only $\sim 40 \mathrm{km\,s^{-1}}$, similar to the LAMOST resolution, TODCOR could detect the faint secondary.

\section{A Method to the \emph{MESS}: Philosophy and Workflow}
\label{sec:mess}

\emph{MESS} considers a set of $M$ observed spectra of a target and evaluates three competing hypotheses:
\begin{itemize}
    \item $\mathcal{S}1$---The target is a single star. 
  All spectra are modeled by a single template with a \emph{single} shared RV.    
    \item $\mathcal{SB}1$---The target is a binary. All spectra are modeled by one stellar spectrum with an RV that varies by epoch.
    \item $\mathcal{SB}2$---The target is a binary; two stellar components are detected in the observed spectra, with epoch-dependent primary and secondary RV shifts.
\end{itemize}

The core of the algorithm is a search over a \textit{continuous} parameter space, generated by interpolating a discrete grid of synthetic templates (characterized by temperature, surface gravity, metallicity, and $V\sin{i}$) to fit the observed spectra. Our aim is not to determine the stellar parameters themselves, but to identify a template (or template pair) that robustly captures the right spectral features and yields a consistent fit across all epochs.
For any proposed set of stellar properties, we generate a synthetic template by interpolating a high-resolution  library of theoretical spectra
(We use \textsc{PHOENIX} by default; \citealt{husser2013}, but this is interchangeable). 
The template is broadened by the proposed $V\sin{i}$, using the standard rotation kernel \citep[e.g.][]{Gray2005}. 

The search is performed via the one-dimensional cross-correlation function \citep[CCF;][]{tonrydavis1979} for the $\mathcal{S}1$ and $\mathcal{SB}1$ models and with TODCOR for the $\mathcal{SB}2$ case. 
For each of the three models,
\emph{MESS} chooses the template (or two templates) that maximizes the weighted mean of the peak correlations \emph{squared} across epochs, with weights reflecting the information content of each epoch. 

This is done by assigning each epoch $m$ a weight $w_m$ that reflects both its Signal-to-Noise Ratio (SNR) and the information
content of the rectified spectrum (Faigler \& Mazeh, to be published):
%
\begin{equation}
w_m \;\equiv\; {\rm SNR}_m^2\cdot{\rm Var}(f_m)\,,
\qquad {\rm and} \qquad
\bar w \equiv \sum_{m=1}^{M} w_m\, .
\label{eq:weights}
\end{equation}
%
For a given model and a specific choice of template parameters, let
$R_{\alpha,m}^\star$ (or $R_{m}^\star$ for $\mathcal{S}1/\mathcal{SB}1$ models) denote the \emph{peak} value of the adopted correlation surface at
epoch $m$ (for $\mathcal{SB}2$, the maximum of $R_\alpha$; for $\mathcal{S}1/\mathcal{SB}1$,
the peak of the 1D-CCF). We aggregate the per-epoch peak strengths into
%
\begin{equation}
S^2 \;\equiv\; \frac{1}{\bar w}\sum_{m=1}^{M} w_m\,\bigl(R_{\alpha,m}^\star\bigr)^2 
\quad {\rm or} \quad
S^2 \;\equiv\; \frac{1}{\bar w}\sum_{m=1}^{M} w_m\,\bigl(R_{m}^\star\bigr)^2 \,,
\label{eq:Sdef}
\end{equation}
%
with $S^2\!\in[0,1]$ by construction (the higher the better). This single score is used both
to rank template candidates and—together with an effective sample size $n_{\rm eff}$—for
model selection (see App.~\ref{app:neff2}).

\subsection{Model selection}
\label{sec:multiepoch}

To decide between the three models, we balance goodness of fit with model simplicity using BIC \citep[]{Schwarz1978} with an \emph{effective} spectral sample size (App.~\ref{app:neff2}).


\noindent The number of parameters, $k$, for each of the three models is:
\begin{itemize}
  \item $k_{\mathcal{S}1}=4+1$.
As all epochs are modeled by the \emph{same} stellar spectrum and a \emph{single} shared RV, we are left with four template parameters, $\{T_{\rm eff},\log g, Z, V\sin{i}\}$, and one shared RV. 

  \item $k_{\mathcal{SB}1}=4+M$.
  The RV is allowed to differ for each of the $M$ epochs.

  \item $k_{\mathcal{SB}2}=8+2M$.
  We now need eight parameters to characterize the two templates, assuming the same metallicity for both components, and a single flux-ratio parameter held fixed across epochs. Two RVs for the primary and secondary for each of the $M$ spectra add another $2M$ parameters. 
\end{itemize}

For each model we derive 
%
\begin{equation}
    \mathrm{BIC} = n_{\rm eff}\,\ln(1-S^2) + k\,\ln n_{\rm eff} \,,
\end{equation}
where the correlation-based fit statistic \(S^2\) aggregates the per-epoch peak correlations (Eq.~\ref{eq:Sdef}) and $k$ is defined above. Here $n_{\rm eff}$ (App.~\ref{app:neff2}, Eq.~\ref{app:eq:neff}) represents the number of independent data points in the system's spectra, calculated via their autocorrelations (see, e.g., \citealt{Bartlett1946,Wilks2011,VonStorchZwiers1999}). For example, while each spectrum in our LAMOST MRS sample contains \(>4000\) pixels, the characteristic \(n_{\rm eff}\) is only \(\sim20\!-\!500\), governed primarily by temperature and \(V\sin{i}\).

The model with the smallest BIC is our \emph{raw} choice. We then apply a short list of astrophysics-motivated overrides for high-confidence edge cases (App.~\ref{app:rules}):
\begin{itemize}[leftmargin=*,itemsep=0.25ex]
\item \textbf{Promote to \(\mathcal{SB}2\):} When the \cite{Wilson1941} linear fit 
(using both-axes uncertainties \citep[see][]{York2004}) displays a significant negative slope (i.e. an astrophysically consistent mass ratio), the estimated RV amplitudes of \emph{both} components exceed the instrument floor, and the phase coverage projected on the Wilson fit is not pathological (App.~\ref{app:wilson}, Eq.~\ref{app:eq:gapPval}). 
\item \textbf{Demote \(\mathcal{SB}2 \to \mathcal{SB}1\):} When the Wilson-plot mass-ratio significance is low or phase-coverage diagnostics are adverse.
\item \textbf{Promote \(\mathcal{S}1 \to \mathcal{SB}1\):} When a single-component model is preferred by BIC but the single-lined RV amplitude proxy is clearly non-zero and exceeds an instrument-dependent threshold.
\item \textbf{Demote \(\mathcal{SB}1 \to \mathcal{S}1\):} When the single-lined RV amplitude proxy falls below the instrument-dependent floor or fails basic robustness checks.
\end{itemize}
These checks guard against false-positive \(\mathcal{SB}2\) assignment due to phase clustering or RV-amplitude noise amplification. The significance of the Wilson-plot fit is another robust tool for identifying $\mathcal{SB}2$ systems without an exhaustive period search. 

The numerical values of the overriding rules used here (App.~\ref{app:simproto}) are currently optimized for the LAMOST telescope, and may vary by instrument owing to telescope-specific systematic errors, spectral bandpass, and related detector properties.

\section{Validation via Simulation}
\label{sec:validation}

To validate our model selection using BIC and the overriding decision scheme, we simulated a sample of $1500$ $\mathcal{S}1$/$\mathcal{SB}1$/$\mathcal{SB}2$ systems with LAMOST MRS-like red-arm spectra, with similar resolving power, pixel sampling, observing epochs, and noise characteristics.

\subsection{Monte Carlo simulation}

The simulation draws binaries from empirically informed distributions where possible, and otherwise from broad priors to cover the key parameters \(\alpha, q, K, e, T_1,\) and \(T_2\). We generated $1500$ synthetic stellar systems, each with \(10\text{--}20\) epochal spectra to match LAMOST observing cadence and instrumental characteristics, reflecting the typical coverage in the LAMOST DR10 multi-epoch dataset where systems have up to 45 epochs but most have fewer than 20; see App.~\ref{app:simproto}.

 For $\mathcal{SB}2$ cases, we generated both components 
from a PHOENIX template grid and drew orbital elements before projecting to radial-velocity time series; for $\mathcal{SB}1$ we suppressed the secondary flux, and for $\mathcal{S}1$ we enforced a constant RV. The forward model includes rotational broadening, Doppler shifts on a log-$\lambda$ grid, and Poisson noise leading to a desired SNR value. 
 
The full \emph{MESS} pipeline (template optimization, TODCOR/1D-CCF, multi-epoch scoring, and BIC-based model selection with physics-motivated overrides) was then run end-to-end on the simulated spectra sets.
Performance was summarized with a row-normalized confusion matrix and by reporting overall accuracy and class-wise precision/recall in table~\ref{tab:confusion_sim} and in figure~\ref{fig:alphavsk}.

\subsection{Key findings}

\begin{figure}
  \centering
  \hspace*{0\columnwidth}
\includegraphics[width=1.0105\columnwidth]{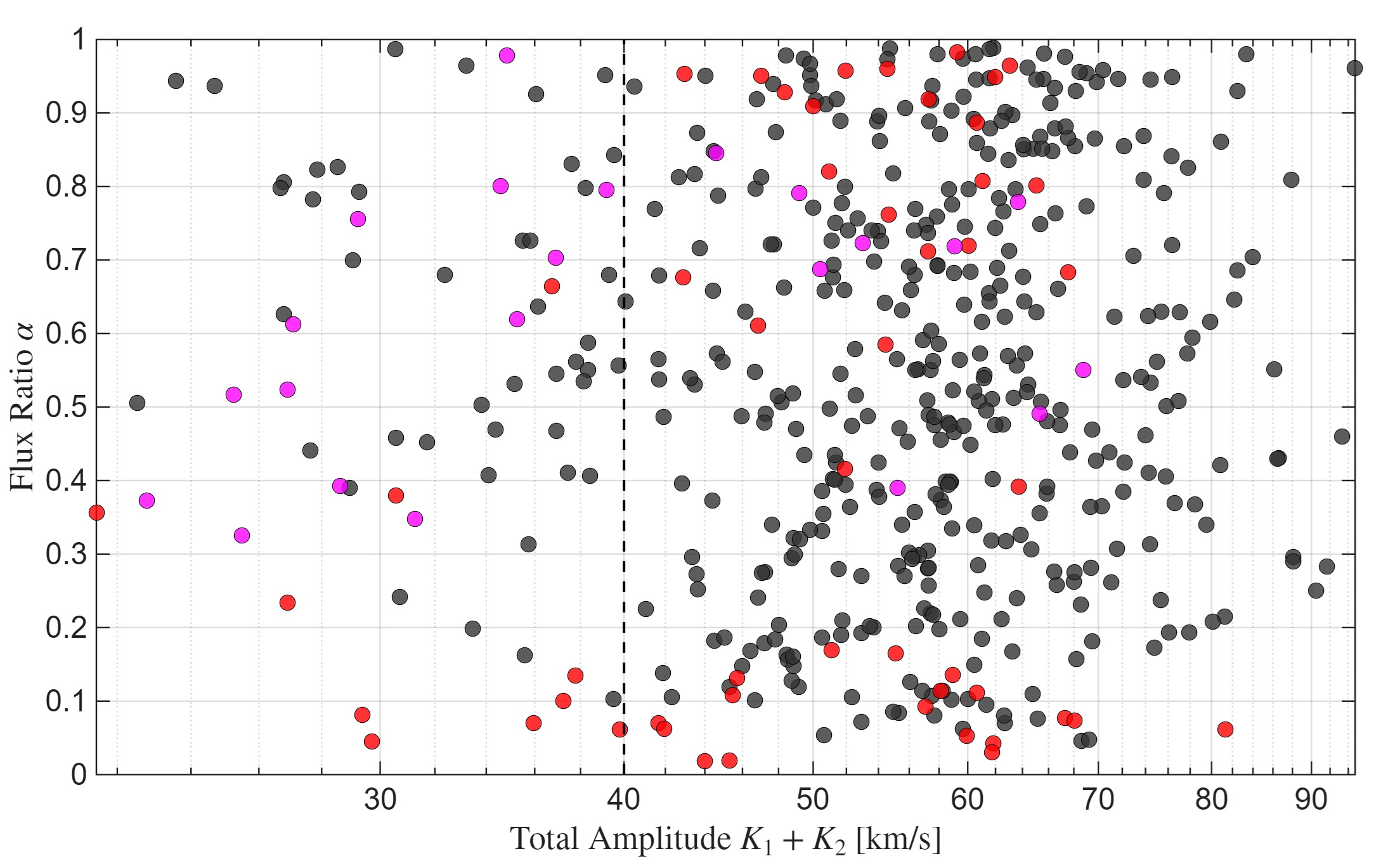}
  \caption{\emph{MESS} results of $500$ simulated $\mathcal{SB}2$ systems in $\alpha$\text{--}$(K_1+K_2)$ space (flux ratio vs.\ the sum of the RV semi-amplitudes). Points are colored by their final label: black---correctly classified as $\mathcal{SB}2$; red---misclassified as $\mathcal{SB}1$; magenta---misclassified as $\mathcal{S}1$. Occurrence rates by outcome are summarized in table~\ref{tab:confusion_sim} (row $\mathcal{SB}2$). The simulated set concentrates on \textit{primary} RV semi-amplitudes below the instrumental resolution; the dashed vertical line marks the LAMOST MRS resolution, $\simeq 40\,\mathrm{km\,s^{-1}}$ ($R\!\approx\!7500$).}
  \label{fig:alphavsk}
\end{figure}

To stress-test \emph{MESS} in its most challenging regimes, we deliberately sampled the parameter space toward low SNR, a limited number of spectra, and small primary RV semi-amplitudes—i.e., predominantly below the instrument's resolution scale (Table~\ref{tab:sim_dists}).
At \(\mathrm{SNR}=50\) and resolving power \(R\simeq 7500\), consistent with the red arm of the LAMOST MRS multi-epoch catalog where most spectra achieve \(\mathrm{SNR}> 50\),
\emph{MESS} cleanly separates \(\mathcal{S}1/\mathcal{SB}/\mathcal{SB}2\) across a broad range of flux ratios, rotational broadening, and orbital phases. The results indicate high fidelity for all three classes, with \(\mathcal{SB}2\) recall remaining strong down to \(\alpha \sim 0.1\) for typical phase coverages. In total, \(79\) of \(1500\) systems (\(\sim5\%\)) were misclassified: 2 \(\mathcal{S}1\) systems were labeled \(\mathcal{SB}2\), 73 \(\mathcal{SB}2\) systems were labeled either \(\mathcal{SB}1\) or \(\mathcal{S}1\) and 4 \(\mathcal{SB}1\) systems were labeled \(\mathcal{S}1\).
Figure~\ref{fig:alphavsk} shows the $500$ simulated $\mathcal{SB}2$ systems in the $(\alpha,K_1+K_2)$ plane (flux ratio versus the sum of RV semi-amplitudes), with points color-coded by classification outcome. Overall, \emph{MESS} performs very well. We observe a clear dependence on the total semi-amplitude: for $K_1+K_2 \lesssim 70\,\mathrm{km\,s^{-1}}$ the success rate declines steadily, consistent with the simulated spectral resolution of $\sim 40\,\mathrm{km\,s^{-1}}$.

Furthermore, the dependence on flux ratio is mild. Misclassifications concentrate in two regimes: (i) at large $\alpha$, when the epoch-by-epoch RV separation is small and the two spectra blend; and (ii) at small $\alpha$, where the secondary is too faint to be reliably detected.

In short, the combination of (i) multi-epoch, correlation-based scoring, (ii) modified information criterion using an effective-sample-size, and (iii) simple astrophysical overrides delivers robust classification at the LAMOST survey scales.

\begin{table}
\centering
\caption{Row-normalized confusion matrix for $1500$ simulated systems ($500$ in each of the $\mathcal{S}1$, $\mathcal{SB}1$, and $\mathcal{SB}2$ classes), all with ${\rm SNR}=50$ and RV semi-amplitudes mostly below, or at most comparable to, the instrumental resolution. The distributions of orbital and stellar parameters are described in App.~\ref{app:simproto}. The simulated PSF, wavelength coverage, and spectral resolution are matched to LAMOST MRS. Conditioning on systems classified as $\mathcal{SB}2$ by \emph{MESS}, only 2 out of 429 were not an intrinsically $\mathcal{SB}2$ system.}

\label{tab:confusion_sim}
\setlength{\tabcolsep}{6pt}
\renewcommand{\arraystretch}{1.2}
\scalebox{1}{%
\begin{tabular}{lccc}
\toprule
\multirow{2}{*}{Truth} & \multicolumn{3}{c}{Selected Model} \\
\cmidrule(lr){2-4}
 & $\mathcal{S}1$ & $\mathcal{SB}1$ & $\mathcal{SB}2$ \\
\midrule
$\mathcal{S}1$   & 99.6\% (498)  & 0.0\% (0) & 0.4\% (2) \\
$\mathcal{SB}1$  & 0.8\% (4)  & 99.2\% (496) & 0.0\% (0) \\
$\mathcal{SB}2$  & 4.4\% (22)  & 10.2\% (51) & 85.4\% (427) \\
\bottomrule
\end{tabular}%
}
\end{table}

\begin{flushleft}
\footnotesize Notes: Entries are percentages with counts in parentheses; each row sums to 100\% (500).
\end{flushleft}

\section{Examples of \emph{MESS} solutions for LAMOST MRS Multi-Epoch systems}
\label{sec:examples}

In this section we use the released LAMOST DR10 red-arm spectra to illustrate \emph{MESS} performance. 
The decision to use the LAMOST red-arm analysis is based on the assumption that the flux ratio in the red band is more favorable for the detection of faint secondaries. We plan to use both arms when applying \emph{MESS} to the entire LAMOST database. 

We use \emph{co-added} spectra, each 
combining several individual exposures, whose single-exposure integration times are typically in the 10–20 min range in LAMOST-MRS operations.
For our tests, we restricted the sample to systems with at least ten epochs, ensuring sufficient phase coverage to exploit the main strengths of \emph{MESS}—in particular, multi-epoch template optimization, BIC-based model selection between $\mathcal{S}1/\mathcal{SB}1/\mathcal{SB}2$, and Wilson-plot mass-ratio inference.

LAMOST radial-velocity zero-point (RVZP) systematics---varying with spectrograph, time, and observing conditions---have been characterized and corrected using RV standards and sky-line anchors in recent studies \citep[e.g.,][]{Zhang2021,Wang2019}. In this preliminary LAMOST showcase, we do not apply any RVZP calibration, and therefore an additional RV variability at the $\sim1-5~\mathrm{km\,s^{-1}}$ level may remain in our measurements.  
A more detailed treatment of the RVZP will be presented in our forthcoming LAMOST catalog paper.
We also did not convert the observation times from MJD to HJD in these examples, since the corresponding timing offsets are negligible for the purposes of demonstrating the algorithm.

The orbital solutions presented in this section are obtained with our own RV–orbit solver: candidate periods are first selected with a $1/\chi^2$ periodogram of $(\rm{v}_1 - \rm{v}_2)^2$ that is robust to velocity-component interchange, and the final Keplerian elements are then derived from a joint, swap-aware fit to both velocity curves followed by a Nelder--Mead refinement. Parameter uncertainties are computed from the local curvature of the $\chi^2$ surface around the best-fit solution.

Running the full pipeline on a compute node with 96 physical cores at $2.3\,$GHz in parallel, the typical wall-clock runtime for a system with $\sim 15\,$ epochs is $\sim 6$ seconds, making the approach practical for survey-scale applications.

\subsection{Example I: disentangling J1145---A low-flux-ratio \texorpdfstring{$\mathcal{SB}2$}{SB2}}
\label{example}

\begin{figure}
  \centering
  \hspace*{0cm}
  \includegraphics[width=0.61\columnwidth]{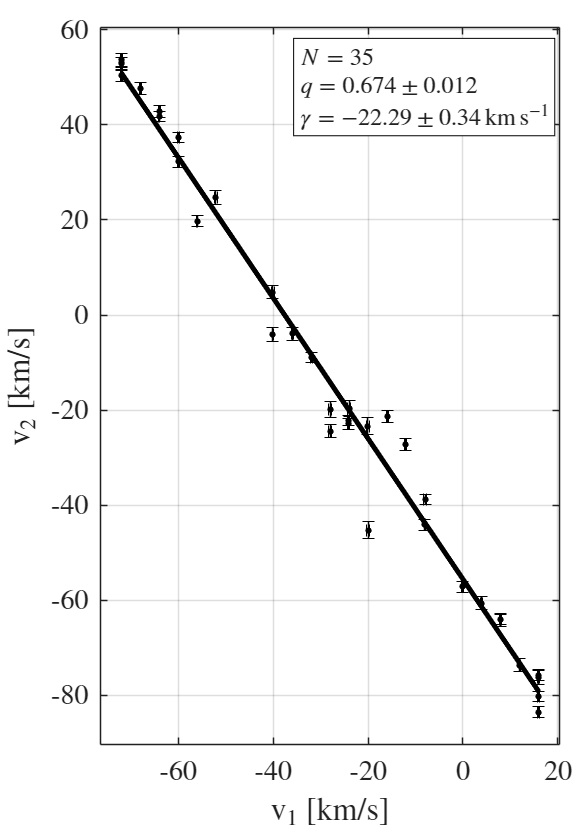}
  \caption{Wilson plot for J1145.
  Points show per-epoch radial-velocity pairs $(\rm{v}_{1,m},\rm{v}_{2,m})$ for $35$ observed epochs; the solid line is the best linear fit, from which the mass ratio $q$ and the systemic velocity $\gamma$ are derived.
  }
  \label{fig:case_wilson}
\end{figure}

\begin{figure*}
    \centering
    \includegraphics[width=1\textwidth]{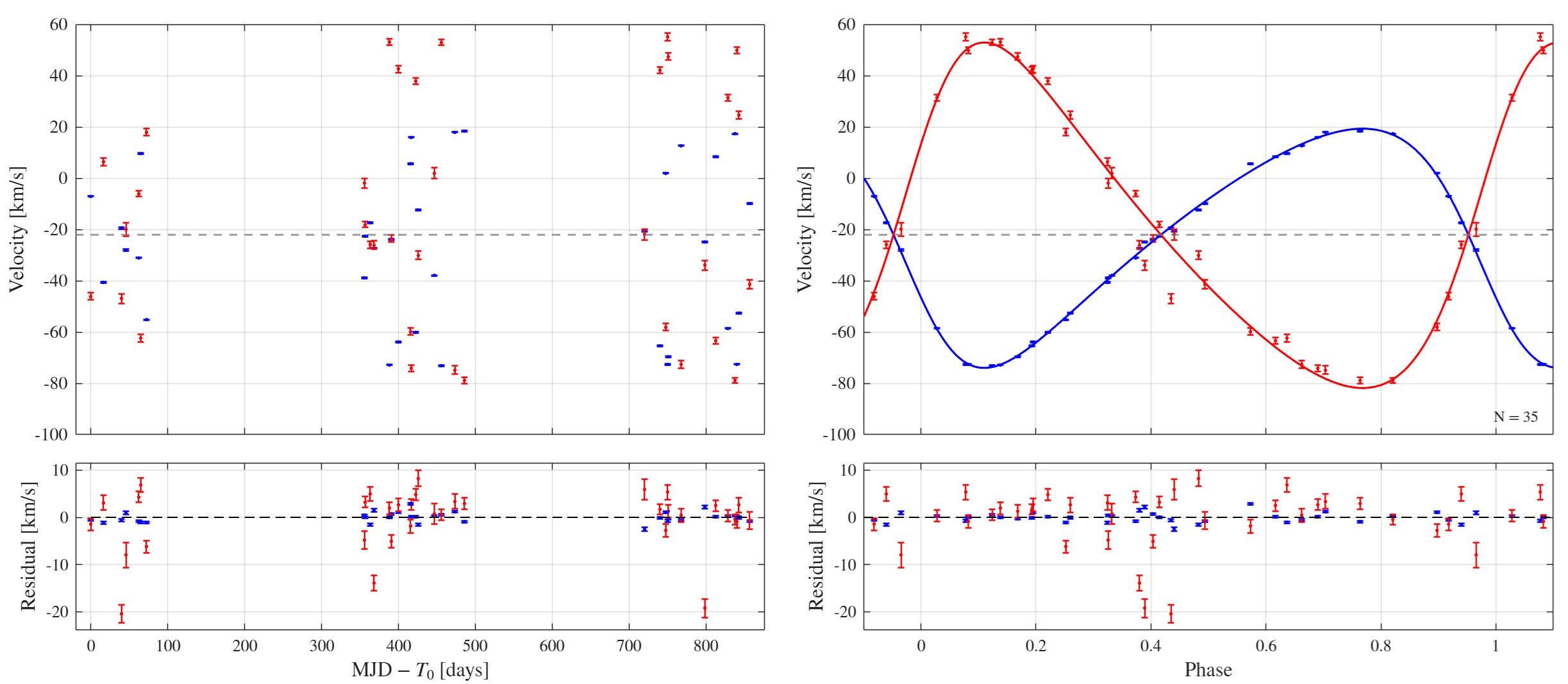}
    \caption{$\mathcal{SB}2$ Keplerian solution for J1145, based on 35 \emph{MESS} RVs, with reference epoch $T_0 = 57016.7$ days.
    Upper-left: primary (blue) and secondary (red) velocities versus ${\rm MJD}-T_0$. Note the time gaps in the data, due to LAMOST observational constraints.}
    Upper-right: the same data, with best-fitting Keplerian curves, phase-folded to the fitted period.
    Lower panels: residuals in time (left) and phase (right).
    \label{fig:case_solution}
\end{figure*}

Figure~\ref{fig:case_wilson} shows the Wilson relation for J1145, the system for which we analyzed a single spectrum in section \ref{sec:todcor:example}, specifically because of its challenging, weak secondary component.
To complete the analysis of J1145, we present in figure~\ref{fig:case_solution} the full Keplerian solution of the system.
The system's best-fit orbital elements and derived parameters are listed in Table~\ref{tab:J1145_orbit}. Throughout this work, we report template temperatures rounded to the nearest integer, while $\log{g}$, $Z$, and $V\sin{i}$ are rounded to the first decimal place.
We emphasize that the 1D analysis would completely miss the secondary component.

\begin{table}
\centering
\caption{LAMOST J114511.45+341926.0 (J1145): $\mathcal{SB}2$ full orbital elements from \emph{MESS}. Wilson-plot estimates are shown alongside the orbital fit. Uncertainties are $1\sigma$. The selected template parameters for this system are $T_{\rm 1}=5747\,\mathrm{K}$, $T_{\rm 2}=4578\,\mathrm{K}$, $\log g_{\rm 1}=3.2$, $\log g_{\rm 2}=3.6$, $Z=-0.5$, $V\sin{i}_{\rm 1}=1.5\,\mathrm{km\,s^{-1}}$ and $V\sin{i}_{\rm 2}=14.1\,\mathrm{km\,s^{-1}}$.}
\label{tab:J1145_orbit}
\setlength{\tabcolsep}{3pt}
\renewcommand{\arraystretch}{1.12}
\small
\newcolumntype{Y}{>{\centering\arraybackslash}X}

\begin{tabularx}{\columnwidth}{@{}l Y Y@{}}
\toprule
 & \textbf{Orbital Solution} & \textbf{Wilson fit} \\
\midrule
$\alpha$ & $0.1064 \pm 0.0001$ & -- \\
\midrule
$P$ (days) & $11.33429 \pm 0.00035$ & -- \\
$e$ & $0.271 \pm 0.005$ & -- \\
$\omega$ (deg) & $114.9 \pm 1.3$ & -- \\
$K_1$ (km\,s$^{-1}$) & $45.6 \pm 0.3$ & -- \\
$K_2$ (km\,s$^{-1}$) & $67.7 \pm 1.5$ & -- \\
$\gamma$ (km\,s$^{-1}$) & $-22.5 \pm 0.2$ & $-22.3 \pm 0.3$ \\
$q$ & $0.673 \pm 0.0151$ & $0.674 \pm 0.012$ \\
\bottomrule
\end{tabularx}

\par\vspace{0.4ex}
\noindent\begin{minipage}{\columnwidth}
\small
\emph{Note. $\alpha$ is measured in the LAMOST~MRS red arm (6300–6800\,\AA).}
\end{minipage}
\end{table}

\subsection{Example II: S1113---A known \texorpdfstring{$\mathcal{SB}2$}{SB2}: comparison with a previous study}
\label{sec:s1113}

Another system we encountered in the LAMOST analysis is 
S1113,\footnote{LAMOST J085125.29+120256.4, Gaia ID 604972089540120832} a short-period ($P\simeq2.823$\,d) double-lined binary in the open cluster M67 with a high-quality TODCOR analysis and orbit by \citet{Mathieu2003}. This binary was identified as an {$\mathcal{SB}2$} in the deep-learning–based LAMOST search by \citet{Li2021_LAMOST_DL_TL} and also appears among the LAMOST-MRS {$\mathcal{SB}2$} orbital solutions of \citet{Guo2025_LAMOSTSB2}.
 Its strong phase-dependent line blending and well-determined literature orbit make it an excellent test case for our algorithm.

\begin{figure}
    \centering
    \includegraphics[width=0.85\columnwidth]{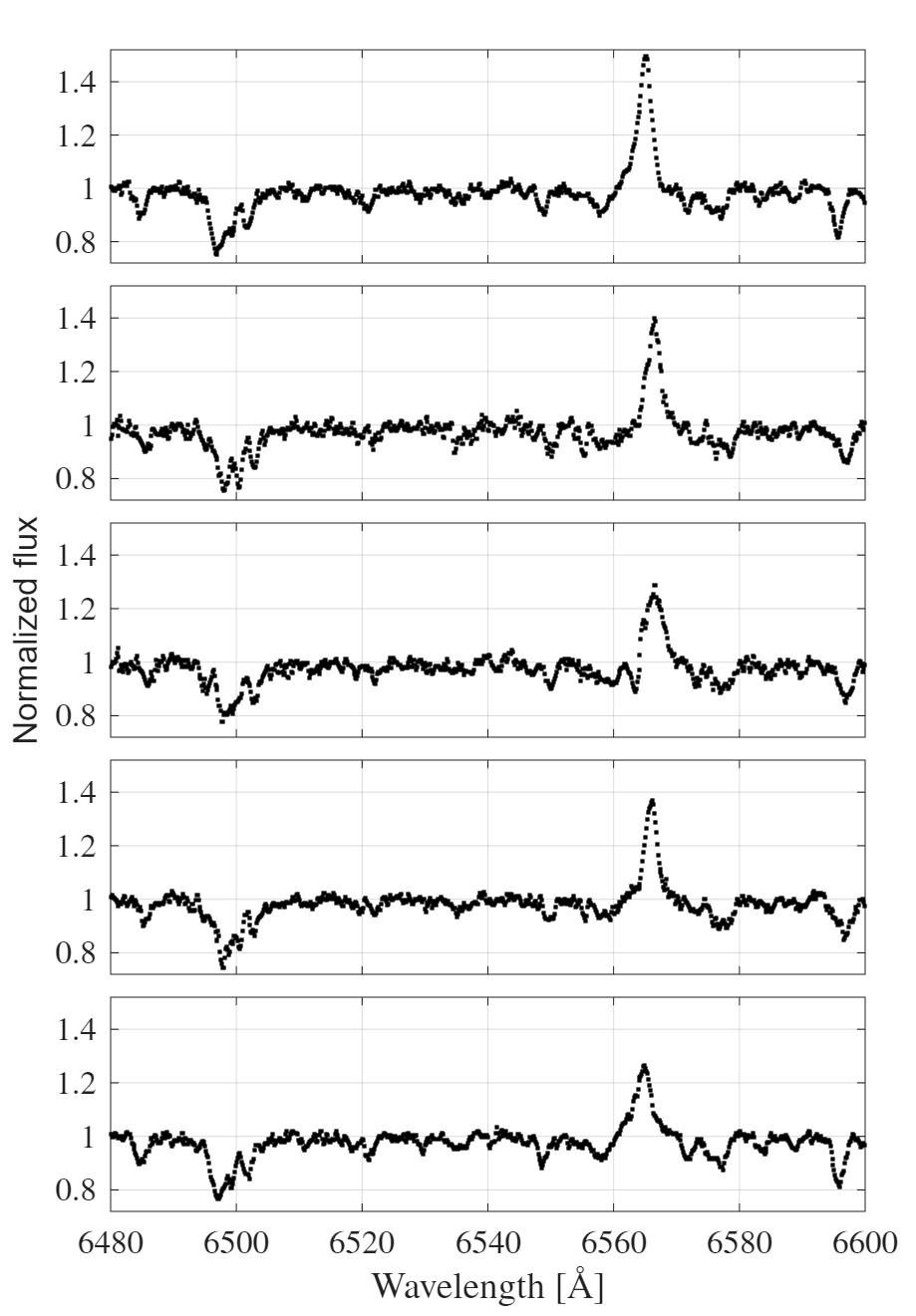}
    \caption{Five representative LAMOST-MRS spectra of S1113 in the H$_{\alpha}$ region, highlighting a narrow, variable, emission component near 6563\,\AA{}. See text for more details.}
    \label{fig:emissionSB2}
\end{figure}

\begin{figure*}
    \centering
    \includegraphics[width=1\textwidth]{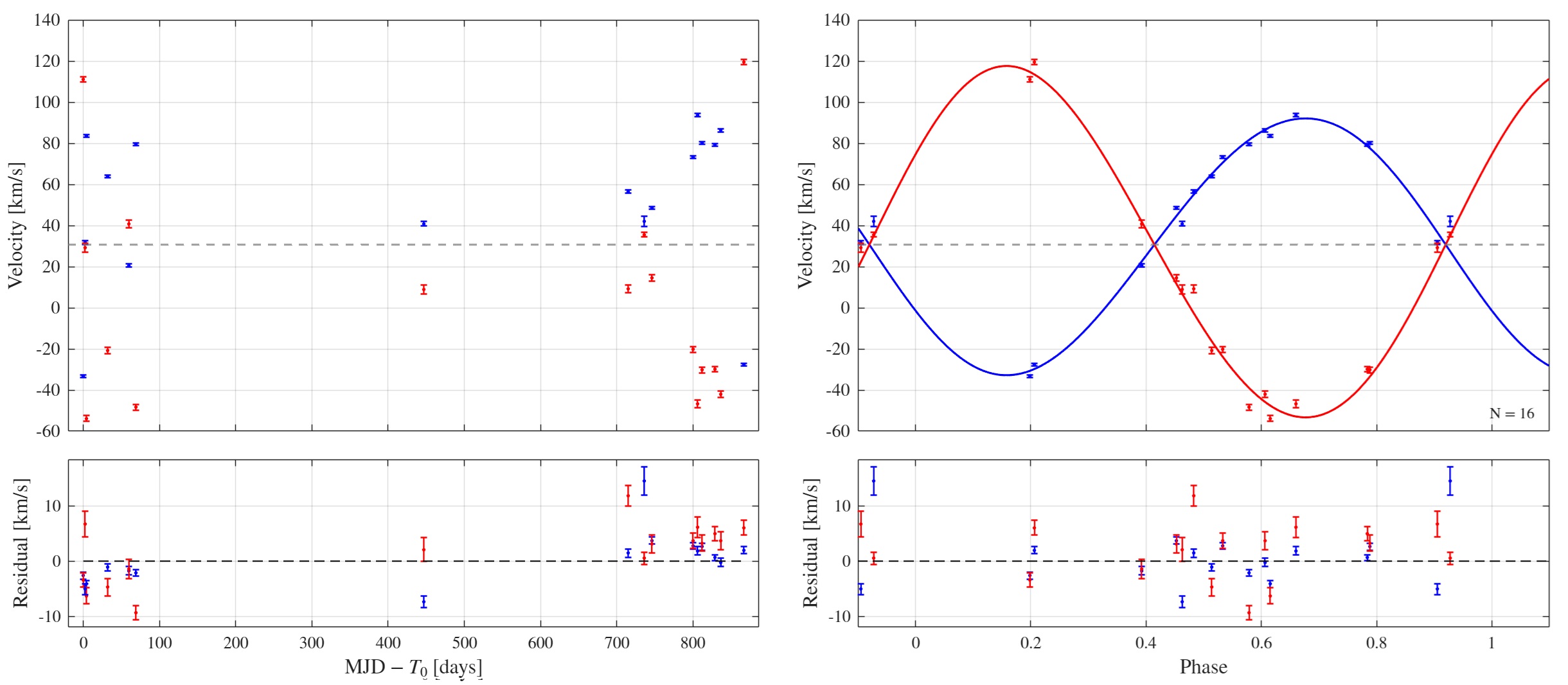}
    \caption{$\mathcal{SB}2$ Keplerian solution for S1113 (M67), based on 16 \emph{MESS} RVs, with reference epoch $T_0 = 56606.7$ days. Plotting conventions follow figure~\ref{fig:case_solution}: primary (blue) and secondary (red) velocities versus ${\rm MJD}-T_0$ (top-left), phase-folded data and solution (top-right), and residuals in time and phase (bottom panels). 
    One may see a long-term trend in the time-domain residuals, indicative of a possible third body in the system.}
    \label{fig:s1113_solution}
\end{figure*}

We note that S1113 shows a narrow H$_{\alpha}$ emission component (Figure~\ref{fig:emissionSB2}). While such emission can in principle bias template matching, the resulting {$\mathcal{SB}2$} solution remains robust. This robustness is supported by Figure~\ref{fig:s1113_solution}, which presents our full orbital solution using the \emph{MESS}-derived RVs, and stems from two factors: (i) the fit is constrained by abundant spectral information well beyond the H${\alpha}$ feature, so the narrow emission contributes only marginally to the global match; and (ii) after continuum-normalizing the spectra—excluding the H${\alpha}$ region from the normalization fit—we apply a normalized-flux cutoff of 1.1 before the subsequent processing, limiting the influence of emission peaks on the solution. We also note a weak, slowly varying trend in the RV residuals as a function of time in Figure~\ref{fig:s1113_solution}, which may indicate the presence of a third component. Such a signal could arise from long-period motion of the inner binary about the barycenter of a hierarchical triple.

Despite different instruments, times of observations, passbands (our 6300–6800\,\AA{} vs.\ the literature’s 5187\,\AA{} window), and template grids, the orbital elements and the mass ratio inferred by \emph{MESS} closely match the independent solution by \citet{Mathieu2003},
as seen in Table~\ref{tab:s1113}. In particular, our mass ratio estimate is within $1\sigma$ of that of \citet{Mathieu2003} and within $2\sigma$ of the value reported by \citet{Guo2025_LAMOSTSB2}.

\begin{figure}
  \centering
  \includegraphics[width=0.71\columnwidth]{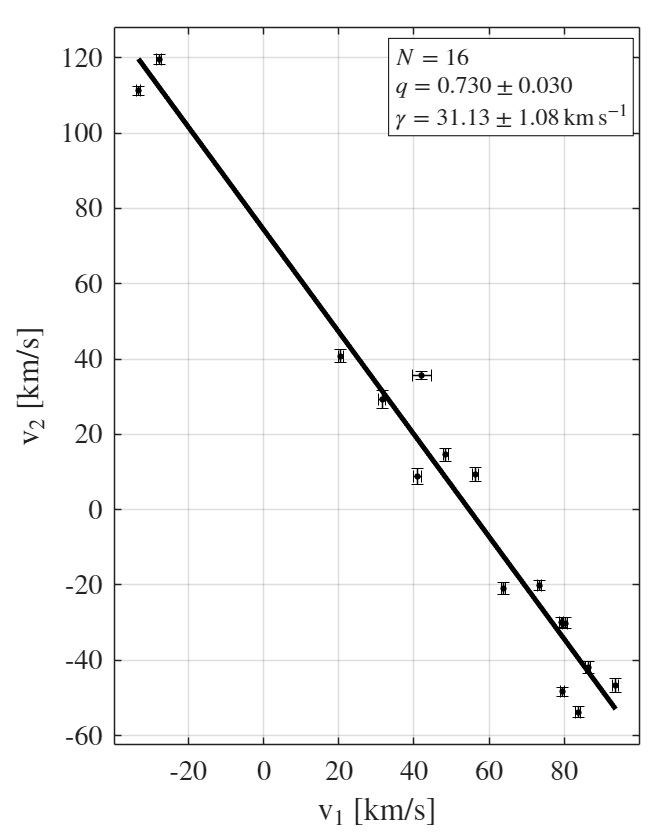}
  \caption{Wilson plot for S1113.
  Points show per-epoch radial-velocity pairs $(\rm{v}_{1,m},\rm{v}_{2,m})$ for $16$ observed epochs; the solid line is the best linear fit, from which the mass ratio $q$ and the systemic velocity $\gamma$ are derived.
  }
  \label{fig:case_wilson2}
\end{figure}

\begin{table}
\centering
\caption{S1113 (M67): Orbital elements from \emph{MESS} compared with the literature \citep{Mathieu2003}. For the mass ratio $q$ and systemic velocity $\gamma$, we report both the Keplerian fit and the Wilson-plot estimates. The selected template parameters for this system are $T_{\rm 1}=4168\,\mathrm{K}$, $T_{\rm 2}=4680\,\mathrm{K}$, $\log g_{\rm 1}=2.9$, $\log g_{\rm 2}=3.8$, $Z=0.2$, $V\sin{i}_{\rm 1}=47.7\,\mathrm{km\,s^{-1}}$ and $V\sin{i}_{\rm 2}=12.3\,\mathrm{km\,s^{-1}}$.}

\label{tab:s1113}
\setlength{\tabcolsep}{2pt}  
\renewcommand{\arraystretch}{1.1}
\small  
\begin{tabularx}{\columnwidth}{@{}l X c c@{}}
\toprule
 & \textbf{This work} & \textbf{\citet{Mathieu2003}} & \textbf{\citet{Guo2025_LAMOSTSB2}} \\
\midrule
$\alpha^{\dagger}$ & $0.208 \pm 0.002$ & $0.35 \pm 0.02$ & $0.34 \pm 0.08$ \\
\midrule
$P$ (days) &
\begin{tabular}[t]{@{}l@{}}$2.822862$\\$\pm 0.000067$\end{tabular} &
\begin{tabular}[t]{@{}c@{}}$2.823094$\\$\pm 0.000014$\end{tabular} &
\begin{tabular}[t]{@{}c@{}}$2.823$\\$\pm 0.539$\end{tabular} \\
$e$ & $0.059 \pm 0.030$ & $0$ (fixed)$^{\ddagger}$ & $0.01 \pm 0.01$ \\
$\omega$ (deg) & $126 \pm 29$ & -- & $-167 \pm 128$ \\
$K_1$ (km\,s$^{-1}$) & $62.4 \pm 1.5$ & $60.6 \pm 0.9$ & $61.25 \pm 0.59$ \\
$K_2$ (km\,s$^{-1}$) & $85.4 \pm 2.6$ & $86.2 \pm 0.6$ & $91.05 \pm 0.47$ \\
$\gamma$ (km\,s$^{-1}$) & $31.4 \pm 0.9$ (orb.) & $33.4 \pm 0.4$ & $31.88 \pm 0.46$ \\
                      & $31.1 \pm 1.1$ (Wil.) &  &  \\
$q$ & $0.73 \pm 0.03$ (orb.) & $0.703 \pm 0.012$ & $0.673 \pm 0.007$ \\
    & $0.73 \pm 0.03$ (Wil.) &  &  \\
\bottomrule
\end{tabularx}

\par\vspace{0.4ex}
\noindent\begin{minipage}{\columnwidth}
\small
\emph{Notes:}
$^{\dagger}$ The flux ratio $\alpha$ in this work is measured over 6300--6800\,\AA, whereas \citet{Mathieu2003} and \citet{Guo2025_LAMOSTSB2} report $\alpha$ at 5187\,\AA, and $\alpha$ at \textasciitilde5000\,\AA, respectively; such bandpass and template-grid differences can produce modest offsets between the reported values.\\
$^{\ddagger}$ A circular solution ($e{=}0$) was adopted as the fitted $e$ was not significantly different from zero: $e=0.022\pm0.010$.
\end{minipage}
\end{table}

\subsection{Example III: J0751---A known \texorpdfstring{$\mathcal{SB}1$}{SB1}: comparison with a previous study}
\label{sec:secJ0751}

\begin{figure*}
    \centering
    \includegraphics[width=1\textwidth]{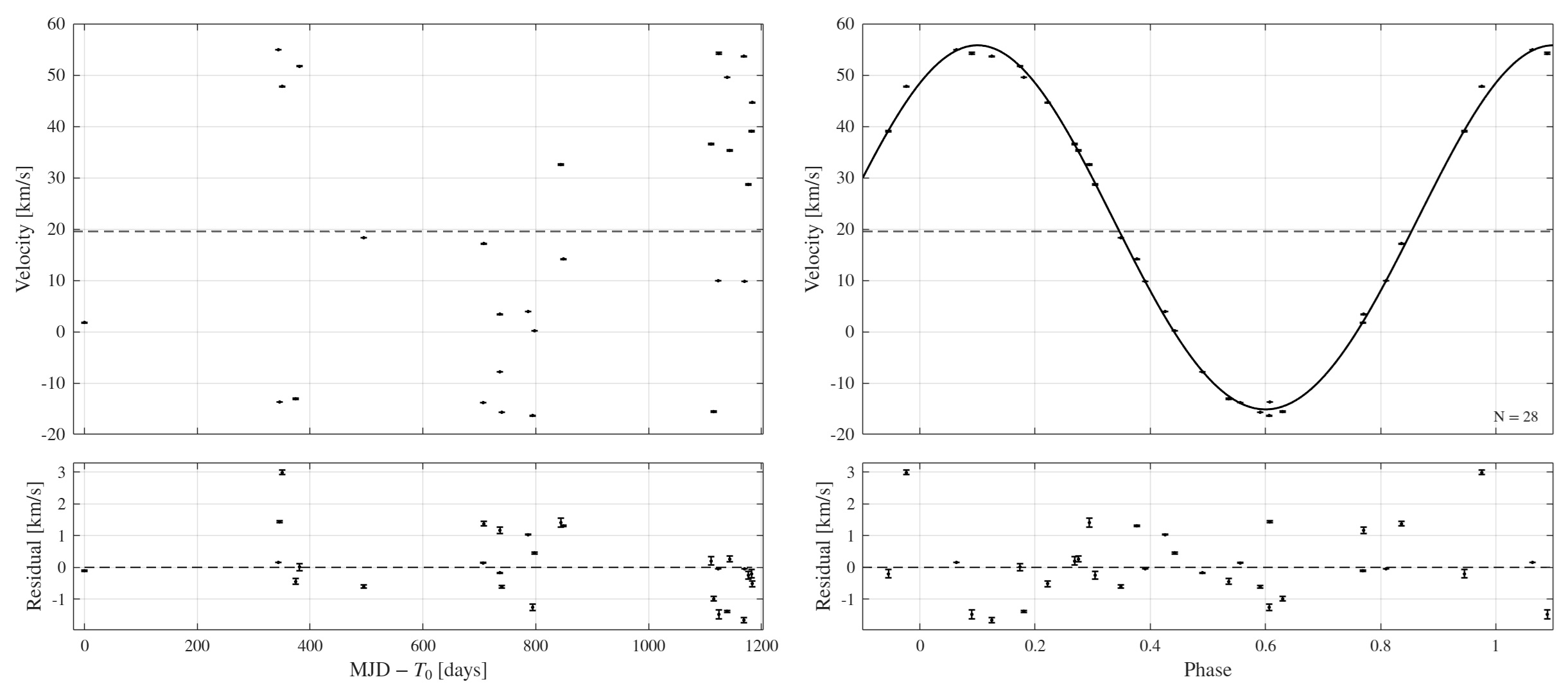}
    \caption{$\mathcal{SB}1$ Keplerian solution for J0751, based on 28 \emph{MESS} RVs, with reference epoch $T_0 = 58029.7$ days. Radial velocities versus ${\rm MJD}-T_0$ (top-left), phase-folded data and solution (top-right), and residuals in time and phase (bottom panels). The orbital elements corresponding to this solution are summarized in table~\ref{tab:tabJ0751} and compared there with the independent solution of \citet{Wu2025PASTVIII}.}
    \label{fig:J0751_solution}
\end{figure*}

Another system identified in our analysis is LAMOST J075154.38+413721.7 (hereafter J0751),\footnote{Gaia ID 920983207515903872} a short-period ($P\simeq3.65$ d) single-lined spectroscopic binary with a recent RV-based orbital solution by \citet{Wu2025PASTVIII} who used \textit{The Joker} code. Figure~\ref{fig:J0751_solution} shows our orbital solution based on the \emph{MESS} RVs, and table~\ref{tab:tabJ0751} presents a side-by-side comparison of the orbital elements obtained in the two studies.

\medskip

\begin{table}
\centering
\caption{Orbital parameters of the single-lined spectroscopic binary LAMOST J075154.38+413721.7 (J0751). 
Values from our \emph{MESS} RV analysis are listed together with the solution reported by \citet{Wu2025PASTVIII}. The selected template parameters for this system are: $T_{\rm 1}=6105 \, \mathrm{K}, \, \log g_{\rm 1}=3.0$, \, $Z=-0.3$, \, $V\sin{i}_{\rm 1}=7.8 \, \mathrm{km\,s^{-1}}$.}
\label{tab:tabJ0751}

\setlength{\tabcolsep}{2pt}  
\renewcommand{\arraystretch}{1.1}
\small  

\begin{tabularx}{0.95\columnwidth}{@{}l >{\centering\arraybackslash}X c@{}}
\toprule
\textbf{} & \textbf{This work} & \textbf{\citet{Wu2025PASTVIII}} \\
\midrule
$P$ (days) & $3.64817 \pm 0.00005$ & $3.65$ \\
$e$ & $0.045 \pm 0.010$ & $0.00$ \\
$\omega$ (deg) & $25.5 \pm 12.7$ & $64.7$ \\
$K_1$ (km\,s$^{-1}$) & $36.1 \pm 0.3$ & $35.1$ \\
$\gamma$ (km\,s$^{-1}$) & $20.5 \pm 0.2$ (orb.) & $21.3$ \\
\bottomrule
\end{tabularx}
\end{table}

\subsection{Example IV: 3 known  \texorpdfstring{$\mathcal{S}1$}{S1} stars}
\label{sec:secJ0622}

\begin{figure*}
    \centering
    \includegraphics[width=\textwidth]{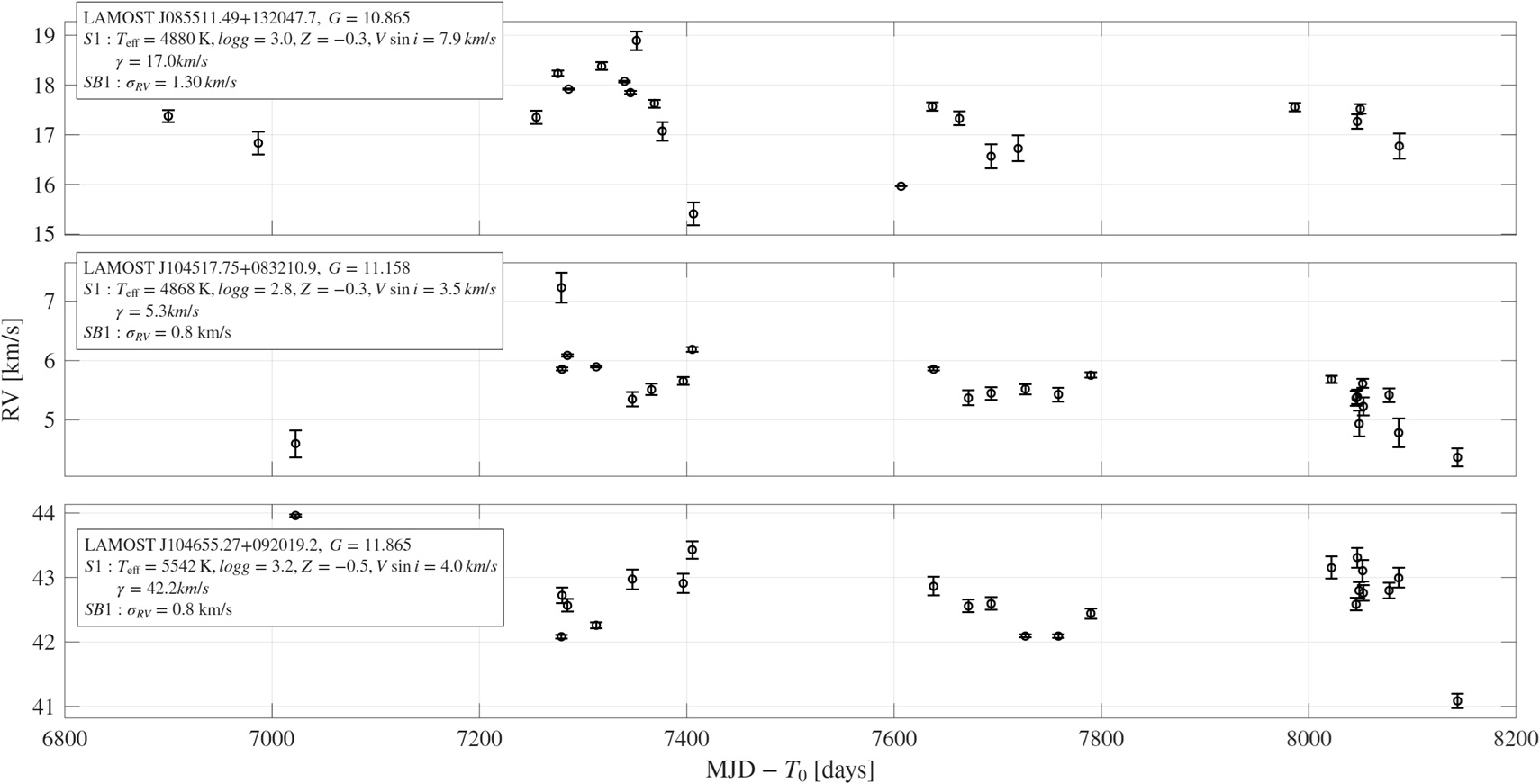}
    \caption{
    {$\mathcal{SB}1$} RVs for three examples of {$\mathcal{S}1$}-classified (single-stars) systems from the RV-stable LAMOST-MRS standard-star sample of \citet{Zhang2021}. From top to bottom, the panels show LAMOST J085511.49+132047.7, LAMOST J104517.75+083210.9, and LAMOST J104655.27+092019.2. Each point represents an epochal RV inferred from the $\mathcal{SB}1$ model of the \emph{MESS} pipeline. The boxed annotations in each panel report the {$\mathcal{S}1$} and {$\mathcal{SB}1$} solution properties. All three stars exhibit RV scatters below $\sim\!1.3\,\mathrm{km\,s^{-1}}$, illustrating the high internal precision of the pipeline for non-variable targets.
    }
    \label{fig:S1_3_examples}
\end{figure*}

\citet{Zhang2021} compiled a catalog of nearly $10^4$ RV-stable standard-star candidates in the LAMOST MRS survey. From this sample, we selected three stars with a substantial number of medium-resolution co-added spectra ($M \geq 25$) and very small reported RV scatter: LAMOST J085511.49+132047.7,\footnote{Gaia ID 605548263696766976} LAMOST J104517.75+083210.9,\footnote{Gaia ID 3868799548307665920} and LAMOST J104655.27+092019.2.\footnote{Gaia ID 3869287211778184064} When processed through the full \emph{MESS} pipeline, all three objects are classified as clean {$\mathcal{S}1$} (single stars), with inferred RV scatters of order $\sigma_{\rm RV} \simeq \mathrm{1\,km\,s^{-1}}$. These systems have effective temperatures in the range $T_{\rm eff}\approx 4800$--$5600$\,K and Gaia magnitudes $G\simeq 11$, and therefore provide representative examples of the internal precision that \emph{MESS} can achieve for non-variable stars observed at LAMOST-MRS. Figure~\ref{fig:S1_3_examples} shows the corresponding RV time series, confirming the absence of significant variability and illustrating the level of scatter expected for bona fide {$\mathcal{S}1$} stars.

%
\section{Discussion}
\label{sec:disc}

In this work, we developed and validated \emph{MESS}---Multi-Epoch Spectroscopic Solver, a TODCOR-based algorithm that analyzes observed stellar spectra and automatically classifies systems as double-lined spectroscopic binaries ($\mathcal{SB}2$), single-lined spectroscopic binaries ($\mathcal{SB}1$), or single stars ($\mathcal{S}1$).
The analysis of $\mathcal{SB}2$ systems is performed by \emph{global} correlation of all observed spectra against a theoretical model composed of two templates interpolated from a synthetic library, yielding the best pair of velocities for each spectrum.
 
\emph{MESS} optimizes the two templates for the observed spectra by searching a \emph{continuous} synthetic-spectra manifold spanning an eight-parameter space—effective temperature, surface gravity, and rotational broadening for each component, together with a common metallicity and the flux ratio $\alpha$. In the present implementation, template evaluation between grid points is obtained by interpolation within the PHOENIX library \citep{husser2013}.
Single-lined spectroscopic binaries ($\mathcal{SB}1$) and single stars ($\mathcal{S}1$) are treated within the same framework by optimizing a single theoretical template for each system.
 
For each system, \emph{MESS} fits three competing models—$\mathcal{S}1$, $\mathcal{SB}1$, and $\mathcal{SB}2$—and selects among them using the Bayesian information criterion (BIC), computed with an effective sample size to account for intra-spectrum correlations. As an auxiliary diagnostic for $\mathcal{SB}2$ candidates, we use the Wilson relation—the linear trend between the per-epoch primary and secondary radial velocities.

We note that a Wilson classification can be misleading: \citet{KovalevWilson24} showed that similar signatures may arise from (i) hierarchical triple system, in which the third component shows a weak RV variability, or (ii) a chance alignment of an unrelated star within the spectroscopic aperture. These cases can produce a nearly constant “secondary" RV and an artificially large mass ratio, mimicking an $\mathcal{SB}2$.
Furthermore, radial-velocity shifts induced by mutual eclipses or star spots can distort the Wilson linear relation. In most cases, these systematics should manifest as coherent structure in the Wilson-residual diagram as a function of observing time—diagnostics we plan to incorporate in our full LAMOST analysis. Indeed, for one of the $\mathcal{SB}2$ systems presented here (S1113), the orbital-residuals show a weak trend suggestive of an additional component.

Nevertheless, although the Wilson approach is not perfect, requiring a significant Wilson fit helps suppress false $\mathcal{SB}2$ positives: when the Wilson regression is weak or inconsistent, the $\mathcal{SB}2$ model is not adopted.

The analogous feature of the $\mathcal{SB}1$ cases would be the complete orbital solution. However, such a solution necessitates a larger number of RV epochs. By contrast, the Wilson line for $\mathcal{SB}2$ systems can be obtained from only a few exposures—fitting just two linear parameters—sufficient to flag $\mathcal{SB}2$ candidates and infer the mass ratio. This economy was the central insight of \citet{Wilson1941}, who derived mass ratios with as few as six spectra. This distinction is crucial in large spectroscopic surveys, where cadence is uneven and most targets have only a handful of observations, often fewer than ten.

The ability of \emph{MESS} to identify $\mathcal{SB}1$ and $\mathcal{SB}2$ systems depends primarily on the stellar spectral type, the signal-to-noise ratio (SNR), and the number of exposures per star, as well as on the spectrograph’s resolution and stability.

\emph{MESS} sensitivity depends on the spectral information contained in the observed bandpass and on the flux ratio between the components. For main-sequence pairs, these factors generally favor the red band—other things being equal. Performance also depends on template fidelity: mismatches between the observed spectra and the synthetic-library theoretical templates degrade both classification and RV precision.

Emission lines can also degrade the quality of template fitting. In \emph{MESS}, the preprocessing removes emission above the continuum level, but cannot eliminate emission that lies within an absorption feature. For example, the S1113 system exhibits clear varying H$_{\alpha}$ emission,
as shown in Figure~\ref{fig:emissionSB2}.
Nevertheless, the $\mathcal{SB}2$ fit for this binary remains of high quality, because the emission feature is weak compared to the overall spectral information in S1113.

In each large survey, cadence, spectral resolution, wavelength coverage, and SNR set the effective detection thresholds for $\mathcal{SB}1$ and $\mathcal{SB}2$ systems as a function of RV semi-amplitude, orbital period, and orbital eccentricity.
Our simulations indicate that our analysis is sensitive to $\mathcal{SB}2$ systems with semi-amplitudes as low as $\sim 10~\mathrm{km\,s^{-1}}$ in the LAMOST dataset. This sensitivity naturally depends on the mass ratio $q$, so for $q \le 1$ the semi-amplitude difference between the primary and secondary must be at least $20~\mathrm{km\,s^{-1}}$.
\cite{kovalevChenHan22,kovalevChen24} reached similar conclusions.

For each $\mathcal{SB}2$ system, the derived mass ratio can be combined with the system’s position on the Gaia color--magnitude diagram (CMD) \citep{GaiaDR3_Summary,GaiaEDR3_Phot}, its combined XP spectrum \citep[e.g.,][]{li25a,li25b}, and available broadband SED photometry \citep[e.g.,][]{andrae23gsp,bayo08vosa} to infer the component masses and stellar parameters. This is especially valuable for modeling binary evolution, particularly when one or both components have evolved away from their zero-age main sequence (ZAMS) position \citep[e.g.,][]{tauris23,marchant25}.

The mass ratio q offers three key advantages: (i) once the inclination is known, it links the orbital solution
to the two dynamical masses, providing benchmark calibrators for stellar models \citep{Torres2010}; (ii) it regulates the \emph{stability and outcome} of mass transfer through critical-$q$ thresholds, separating stable Roche-lobe overflow, thermal-timescale transfer, and common-envelope evolution or merger \citep{Hjellming1987,Soberman1997,Ivanova2013}; and (iii) at the population level, the $q$ distribution constrains \emph{formation and evolutionary pathways}, because it encodes how binaries form and interact across mass and period \citep{Moe2017}.

\emph{MESS} is immediately applicable to current and upcoming large spectroscopic surveys. We are applying it to the recent LAMOST medium-resolution survey release ($\,\sim8$ million spectra). In a selected subsample, we have identified $\sim 2{,}000$ binaries, $25\%$ of which are $\mathcal{SB}2$ systems, with flux ratios as low as $\sim 0.1$ (Nachmani, Faigler \& Mazeh, to be published). Two examples are shown above.
In the present application of MESS to LAMOST, we analyze the co-added spectra, for which the total exposure time is typically on the order of an hour per epoch. Consequently, this setup has limited sensitivity to very short-period binaries and is not optimal for reliably identifying or characterizing such systems.

Recently, \citet{wang25} applied a 1D cross-correlation to the LAMOST DR11 blue-arm spectra and identified nearly \(8000\) \(\mathcal{SB}2\) candidates. By design, \emph{MESS}—which uses 2D cross-correlation and explicitly fits the flux ratio \(\alpha\)—is more sensitive to faint secondaries in the LAMOST red-arm and is therefore expected to recover additional systems missed by 1D searches.

In addition to LAMOST, we are witnessing a surge of large spectroscopic surveys with public data releases, all of which can benefit from applying \emph{MESS}. These include SDSS/APOGEE \citep{Majewski2017APOGEE}, GALAH \citep{DeSilva2015GALAH}, the Gaia–ESO Survey \citep{Randich2022GES}, the DESI Milky Way Survey \citep{Cooper2023DESIMWS}, and 4MOST \citep{dejong2019}; notably, the Gaia RVS project \citep[e.g.,][]{RVS18} is expected to release more than $10^{9}$ spectra in the forthcoming DR4 \citep{Gaiabrown25}.\footnote{\url{https://www.cosmos.esa.int/web/gaia/data-release-4}}

We expect \emph{MESS} to identify many thousands of $\mathcal{SB}2$ systems with measured mass ratios, enabling subsequent determination of component masses and stellar parameters. Large, homogeneous $\mathcal{SB}2$ samples will then yield the binary mass-ratio distribution directly from observations, without reliance on statistical inference or deconvolution \citep*[e.g.,][]{Goldberg2003,Arenou2015_deconvolution,sahar17,Boffin2018_massratioSBs,sahar19}.
This, in turn, will enable studying the 
mass-ratio distribution as a function of orbital parameters, such as period and eccentricity, and stellar mass 
and angular momentum \citep[e.g.,][]{duchene13}---key features in the study of binary formation and evolution. The primordial mass-ratio distribution, in massive stars in particular, is a crucial element in our understanding of the history of the emerging population of black-hole binaries \citep[e.g.,][]{BH3, nagarajan25, GWTC-4.0}.

\section*{Software and Data Availability}

The pipeline is implemented in MATLAB as \texttt{multiCorr}. Synthetic templates were drawn from PHOENIX/ATLAS9 grids. Example scripts and configuration files will be released in a companion repository.

\section*{Acknowledgments}

We deeply thank the referee, M. Kovalev, for many illuminating comments that substantially improved the paper.
We thank the LAMOST and Gaia teams for making their survey data publicly available, and our colleagues at Tel Aviv University, Elad Goldberg and Shay Zucker, for illuminating discussions and brilliant suggestions.

\newpage

\bibliographystyle{mnras}
\bibliography{references}


\appendix
\numberwithin{equation}{section} 
\renewcommand{\thesubsection}{\thesection\arabic{subsection}} 

\makeatletter
\renewcommand{\thesubsection}{\Alph{section}\arabic{subsection}}
\let\OldSection\section
\let\oldsection\section
\renewcommand{\section}[1]{%
  \refstepcounter{section}%
  \setcounter{subsection}{0}%
  \setcounter{equation}{0}%
  \OldSection*{APPENDIX \Alph{section}}%
}
\makeatother

\providecommand{\std}{\mathop{\mathrm{std}}\nolimits}

\label{app:metric}
\begingroup
\setlength{\abovedisplayskip}{8pt}
\setlength{\belowdisplayskip}{8pt}
\setlength{\abovedisplayshortskip}{6pt}
\setlength{\belowdisplayshortskip}{6pt}
\setlength{\textfloatsep}{8pt}
\setlength{\floatsep}{8pt}
\setlength{\intextsep}{7pt}

\makeatletter
\renewcommand{\section}[1]{%
  \refstepcounter{section}%
  \setcounter{subsection}{0}%
  \setcounter{equation}{0}%
  \phantomsection
  \OldSection*{APPENDIX \Alph{section} -- #1}
  \addcontentsline{toc}{section}{Appendix \Alph{section} -- #1}%
  \markboth{Appendix \Alph{section}: #1}{Appendix \Alph{section}: #1}%
}
\makeatother

\section{Symbols and Definitions}
\label{app:symbols}

\setcounter{equation}{0}
\renewcommand{\theequation}{A\arabic{equation}}

\addcontentsline{toc}{section}{Symbols and definitions}

\medskip
\noindent\textit{Grid, data, and basic weights}
\begin{symbollist}
  \item[$N$] Number of pixels per spectrum on the log-$\lambda$ grid.
  \item[$\{\lambda_n\}_{n=1}^{N}$] Wavelength grid, equally spaced in $\ln\lambda$.
  \item[$d v=c\,\Delta\ln\lambda$] Constant Doppler step per pixel; App.~\ref{app:prelim}.
  \item[$c$] Speed of light.
  \item[$M$] Number of epochs/spectra; index $m=1,\ldots,M$.
  \item[$f_m$] Continuum-normalized, mean-subtracted spectrum at epoch $m$.
  \item[${\rm SNR}_m$] Reported signal-to-noise ratio for epoch $m$.
  \item[$w_m$] Epoch weight $\propto {\rm SNR}_m^2\cdot{\rm Var}(f_m)$; $\bar w=\sum_m w_m$ (Eq.~\ref{eq:weights}).
\end{symbollist}

\noindent\textit{Templates, kinematics, and rotation}
\begin{symbollist}
  \item[$g_1,g_2$] Primary/secondary synthetic templates.
  \item[$\rm{v}_{1,m},\rm{v}_{2,m}$] Primary/secondary RVs (km\,s$^{-1}$) at epoch $m$.
  \item[$\alpha$] Monochromatic flux ratio $F_2/F_1$ (epoch-invariant).
  \item[$V\sin{i}$] Projected rotational velocity (km\,s$^{-1}$).
  \item[$\epsilon$] Linear limb-darkening coefficient in the rotational kernel.
  \item[$\mathcal{G}(\rm{v};V\sin{i},\epsilon)$] Rotational broadening kernel; Eq.~\eqref{app:eq:rotkernel}.
\end{symbollist}

\noindent\textit{Correlations and multi-epoch score}
\begin{symbollist}
  \item[$C_1,C_2,C_{12}$] Normalized correlations: $(f,g_1)$, $(f,g_2)$, and $(g_1,g_2)$.
  \item[$R_\alpha(s_1,s_2)$] TODCOR surface for a given flux ratio $\alpha$; Eq.~\eqref{app:eq:Ralpha_recall}.
  \item[$S$] Weighted multi-epoch peak-correlation score; Section~\ref{sec:multiepoch}, Eq.~\ref{eq:Sdef}.
\end{symbollist}

\noindent\textit{Model selection and effective size}
\begin{symbollist}
  \item[$\rho_m(k)$] Lag-$k$ autocorrelation of the observed spectrum in epoch $m$.
  \item[$n_{{\rm eff},m}$] Effective independent samples at epoch $m$; Eq.~\eqref{app:eq:neff}.
  \item[$n_{\rm eff}$] Sum over epochs of $n_{{\rm eff},m}$.
  \item[$k$] Parameter count for a model ($\mathcal{S}1$/$\mathcal{SB}1$/$\mathcal{SB}2$).
  \item[AIC, BIC] Information criteria; Eqs.~\eqref{app:eq:aic}–\eqref{app:eq:bic}.
\end{symbollist}

\noindent\textit{Wilson fit and coverage diagnostics}
\begin{symbollist}
  \item[$a,b$] slope and intercept of Wilson fit (~\ref{app:wilson}).
  \item[$q,\gamma$] Mass ratio and systemic velocity from $(a,b)$; Eq.~\eqref{app:eq:wgamma}.
  \item[$\sigma_q,\sigma_\gamma$] Errors propagated from $(a,b)$; Eqs.~\eqref{app:eq:siggamma}.
  \item[$\delta$] Largest normalized spacing along the projected Wilson line.
  \item[$p_{\rm gap}(\delta;M)$] Probability of observing a gap $\ge\delta$ under uniform coverage by $M$ points; Eq.~\eqref{app:eq:gapPval}.
\end{symbollist}

\noindent\textit{Orbit and derived velocities}
\begin{symbollist}
  \item[$P,e,i,\omega,T_0$] Period, eccentricity, inclination, argument of pericenter, reference epoch.
  \item[$K_1,K_2$] RV semi-amplitudes of the primary/secondary.
  \item[$\widehat K_1,\widehat K_2$] Dispersion-based amplitude estimates; Eq.~\eqref{app:eq:Kproxies}.
  \item[$\hat{\sigma}$] The standard deviation operator.
  \item[$\gamma$] Systemic velocity (also from the Wilson mapping).
\end{symbollist}

\noindent\textit{Likelihood model selection}
\begin{symbollist}
\item[$\boldsymbol{\theta}$] Generic model-parameter vector; one of
    $\boldsymbol{\theta}_{\mathcal{S}1}$, $\boldsymbol{\theta}_{\mathcal{SB}1}$, or
    $\boldsymbol{\theta}_{\mathcal{SB}2}$ (template parameters only; per-epoch velocities are listed separately).
  \item[$\boldsymbol{\theta}_{\mathcal{S}1}$] Single-star (no RV variation) template:
    $\boldsymbol{\theta}_{\mathcal{S}1}=(T,\log g,Z,V\sin{i})$.
  \item[$\boldsymbol{\theta}_{\mathcal{SB}1}$] Single-lined binary template:
    $\boldsymbol{\theta}_{\mathcal{SB}1}=(T,\log g,Z,V\sin{i})$.
  \item[$\boldsymbol{\theta}_{\mathcal{SB}2}$] Double-lined binary template pair:
    $\boldsymbol{\theta}_{\mathcal{SB}2}=(T_1,T_2,\log g_1,\log g_2,Z,V\sin{i}_1,V\sin{i}_2,\alpha)$,
    with a shared metallicity $Z$, projected rotations $(
    V\sin{i})_1$ and $(V\sin{i})_2$ and monochromatic flux ratio $\alpha\!=\!F_2/F_1$.
\end{symbollist}

\section{Core Methods}
\label{app:core}
\addcontentsline{toc}{section}{Appendix B — Core methods: data, correlations, errors, and model selection}
\setcounter{equation}{0}
\renewcommand{\theequation}{B\arabic{equation}}

This section details our data model and weighting, correlation measures (1D-CCF and TODCOR), peak refinement and RV errors, rotational broadening, effective sample size with information criteria, Wilson-plot inference (including coverage diagnostics), the rule-based model-selection checks, and the scale-eliminated likelihood machinery for parameter/flux-ratio uncertainties.

\subsection{Grid, data, and multi-epoch score}
\label{app:prelim}

We work on a logarithmic wavelength grid $\{\lambda_n\}_{n=1}^{N}$ so that a single pixel shift
corresponds to a \emph{constant} Doppler step $d v = c\,\Delta\ln\lambda$, which simplifies 
sub-pixel interpolation and RV error propagation. Each target has $M$ continuum-normalized,
mean-subtracted spectra $\{f_m(\lambda)\}_{m=1}^{M}$.

For given stellar parameters, we draw two synthetic templates $g_1,g_2$ (convolved to the
instrumental resolution and rotationally broadened; App.~\ref{app:rot}). Epoch weights $w_m$
and the multi-epoch score $S^2$ used for optimization and model selection are defined in
section~\ref{sec:multiepoch}, Eqs.~\eqref{eq:weights}–\eqref{eq:Sdef}.

For per-epoch matching we adopt the normalized 1D-CCF for $\mathcal{S}1/\mathcal{SB}1$ and the
TODCOR surface for $\mathcal{SB}2$ exactly as in section~\ref{sec:todcor}. The discrete maximum
provides initial lag(s); subpixel refinement yields per-epoch velocities (App.~\ref{app:refine}).


\subsection{Peak refinement and per-epoch RV errors}
\label{app:refine}

\textbf{Subpixel peak.}
We take the 3×3 neighborhood around the discrete peak of the correlation surface (or curve in 1D). The peak position is refined by fitting simple three-point parabolas along the central row and column (standard quadratic interpolation), yielding sub-pixel offsets in the two lag directions, and the corresponding velocities. In the final polishing, we up-sample the correlation map (using a default multiplier of 10).

\noindent\textbf{Uncertainty estimation.}
Let \(R(\rm{v}_1, \rm{v}_2)\) denote the (normalized) correlation surface evaluated on the lag grid, and let \(Z\) represent the \(3\times3\) neighborhood centered around its discrete maximum (center element \(Z_{2,2}\); rows correspond to \(\rm{v}_1\), columns to \(\rm{v}_2\)).
Let the pixel spacings along the two lag axes be \(\Delta \rm{v}_1\) and \(\Delta \rm{v}_2\) (in velocity units).

\noindent\emph{Local quadratic approximation.}
Assuming \(R\) is sufficiently smooth near the maximum, we approximate it via a second-order Taylor expansion and compute the \(2\times2\) Hessian using centered second finite differences:
\begin{equation}
\begin{aligned}
\kappa_{11} &= \frac{Z_{1,2} - 2Z_{2,2} + Z_{3,2}}{(\Delta \rm{v}_1)^2}, \\
\kappa_{22} &= \frac{Z_{2,1} - 2Z_{2,2} + Z_{2,3}}{(\Delta \rm{v}_2)^2}, \\
\kappa_{12} &= \frac{Z_{3,3} - Z_{3,1} - Z_{1,3} + Z_{1,1}}{4\,\Delta \rm{v}_1 \Delta \rm{v}_2}\,.
\end{aligned}
\label{app:eq:kappas}
\end{equation}
The Hessian is defined as \(H=\begin{psmallmatrix}\kappa_{11}&\kappa_{12}\\\kappa_{12}&\kappa_{22}\end{psmallmatrix}\), \(\hat R\equiv R|_{\text{peak}}\), and the scaling \(\eta=\dfrac{n_{\rm eff}\,\hat R}{1-\hat R^{2}}\), where \(n_{\rm eff}\) is the effective number of independent spectral samples (see section~\ref{app:neff2}). For a valid local maximum (i.e., \(-H\) is positive-definite and well-conditioned), a Gaussian/Laplace approximation to the log-likelihood \(\log \mathcal{L}(\boldsymbol{\theta}) = -\tfrac{n_{\rm eff}}{2}\,\ln\!\bigl(1 - R(\boldsymbol{\theta})^{2}\bigr)\) gives the parameter covariance (in velocity units) as
\(\Sigma \approx [\eta(-H)]^{-1}\) (e.g., \citealp{vanderVaart1998,EfronHinkley1978}).

\noindent\textbf{Noise floor.}
Let \(\sigma_{\rm floor}\) (km\,s\(^{-1}\)) denote an instrument- and SNR-dependent RV noise floor.
We combine it in quadrature with the curvature-based covariance:
\begin{equation}
\Sigma^{\rm tot} \;=\; \Sigma \;+\; \sigma_{\rm floor}^{2}\,I_{2}\,,
\label{app:eq:verr2}
\end{equation}
where $I_{2}$ is the $2\times2$ identity matrix. We then report the \(1\sigma\) uncertainties as
\begin{equation}
\sigma_{\rm{v}_1} \;=\; \sqrt{\Sigma^{\rm tot}_{11}}, \qquad
\sigma_{\rm{v}_2} \;=\; \sqrt{\Sigma^{\rm tot}_{22}}\,.
\end{equation}

\noindent\textbf{Fallback (axis-wise curvature).}
If \(-H\) is not positive-definite or is ill-conditioned, we estimate the per-axis errors from the one-dimensional curvatures (which are negative at a local maximum):
\begin{equation}
\sigma_{\rm{v}_1,{\rm curv}} \;=\; \frac{1}{\sqrt{\eta\,(-\kappa_{11})}}, \qquad
\sigma_{\rm{v}_2,{\rm curv}} \;=\; \frac{1}{\sqrt{\eta\,(-\kappa_{22})}}\,,
\end{equation}
and include the same noise floor as in Eq.~\eqref{app:eq:verr2}:
\begin{equation}
\sigma_{\rm{v}_1} \;=\; \sqrt{\sigma_{\rm{v}_1,{\rm curv}}^{2} + \sigma_{\rm floor}^{2}}, \qquad
\sigma_{\rm{v}_2} \;=\; \sqrt{\sigma_{\rm{v}_2,{\rm curv}}^{2} + \sigma_{\rm floor}^{2}}\,.
\end{equation}

\noindent\textbf{SB1 case.}
For single-lined spectra, we apply the same methodology along the single axis using the corresponding one-dimensional curvature and spacing \(\Delta v\).

\subsection{Rotational broadening on a log-\texorpdfstring{$\lambda$}{lambda} grid}
\label{app:rot}
We adopt the classical rotational broadening kernel with projected rotation $V\sin{i}$ and a fixed linear limb–darkening coefficient \(\epsilon=0.6\) \citep{Gray2005,OroszHauschildt2000}:

\begin{equation}
\label{app:eq:rotkernel}
\begin{aligned}
\mathcal{G}(\rm{v};V\sin{i},\epsilon) &=
\begin{cases}
\dfrac{2(1-\epsilon)\sqrt{1-x^{2}} + \dfrac{\pi}{2}\epsilon(1-x^{2})}{\pi\,V\sin{i}\,(1-\epsilon/3)}\,, & |x|<1\,,\\[6pt]
0, & \text{otherwise}\,,
\end{cases}\\
&\text{where } x \equiv \dfrac{\rm{v}}{V\sin{i}}\,.
\end{aligned}
\end{equation}
On the log-grid we convolve $g$ with $\mathcal{G}$ after mapping the velocity support to an integer pixel span using $d v$; reflection padding prevents edge losses.

\subsection{Effective sample size and information criteria}
\label{app:neff2}
The number of independent data points of an observed spectrum is derived via its autocorrelation \citep[e.g.][]{Bartlett1946,Wilks2011,VonStorchZwiers1999}:
\begin{equation}
n_{{\rm eff},m} \;=\; \frac{N}{\,1 + 2\sum_{k=1}^{L_m} \bigl(1-\tfrac{k}{N}\bigr)\rho_{m}(k)\,} \,,
\label{app:eq:neffm}
\end{equation}
\noindent
\noindent
where $\rho_m(k)$ is the autocorrelation at lag $k$ of the mean-subtracted spectrum at epoch $m$, and $L_m$ is the truncation lag beyond which autocorrelations are treated as negligible (e.g., the first non-positive $\rho_m$, or a Newey--West bandwidth with a Bartlett kernel \citep{NeweyWest1987,NeweyWest1994}). To obtain the overall effective sample size, sum over epochs:
\begin{equation}
n_{\rm eff} \;=\; \sum_{m=1}^{M} n_{{\rm eff},m}\,.
\label{app:eq:neff}
\end{equation}
 Using \(n_{\rm eff}\ln\!\bigl(1-S^2\bigr)\) as our fit statistic, the information criteria balance fit and model complexity via penalty terms—AIC \citep[]{Akaike1974} adds \(2k\) and BIC \citep[]{Schwarz1978} adds \(k\ln n_{\rm eff}\)—so
\begin{align}
\mathrm{AIC} &= n_{\rm eff}\,\ln(1-S^2) + 2k\,, \label{app:eq:aic}\\
\mathrm{BIC} &= n_{\rm eff}\,\ln(1-S^2) + k\ln n_{\rm eff}\,, \label{app:eq:bic}
\end{align}
with $k_{\scriptstyle{\mathcal{\scriptscriptstyle S}1}}{=}5$, $k_{\scriptstyle{\mathcal{\scriptscriptstyle SB}1}}{=}4{+}M$, $k_{\scriptstyle{\mathcal{\scriptscriptstyle SB}2}}{=}8{+}2M$.

\subsection{Wilson-plot inference for \texorpdfstring{$(q,\gamma)$}{(q,gamma)}}
\label{app:wilson}
The Wilson relation regresses $\rm{v}_2$ on $\rm{v}_1$,
\begin{equation}
\rm{v}_2 \;=\; a\,\rm{v}_1\;+\; b,
\end{equation}
with errors in both axes (\cite{York2004}). In a Keplerian binary\\ $a=-K_2/K_1=-1/q<0$, where $K_1$ and $K_2$ are the semi-amplitudes of the components and $q$ is their mass-ratio. From $(a,b)$,
\begin{equation}
q \;=\; -\frac{1}{a}\,, 
\qquad 
\gamma \;=\; \frac{b}{1-a}\,,
\label{app:eq:wgamma}
\end{equation}
%
and the $(a,b)$ covariance gives $\sigma_q$ and $\sigma_\gamma$:
\begin{equation}
\sigma_q \;=\; \frac{\sigma_a}{|a|^{2}}\,, \qquad
\sigma_\gamma^2 \;=\; \frac{\sigma_b^2}{(1-a)^2} + \frac{b^2\,\sigma_a^2}{(1-a)^4} + \frac{2b\,\sigma_{ab}}{(1-a)^3}\,.\label{app:eq:siggamma}
\end{equation}

\textbf{Coverage diagnostic (max–gap test) for model selection.}
Project the measured pairs onto the best-fit Wilson line, sort the projected coordinates, rescale their lengths to $[0,1]$. Let $M$ be the number of projected data points ($=$ number of epochs) and let
$\delta=\max_{1\le i\le M-1}\{u_{(i+1)}-u_{(i)}\}$ be the largest normalized spacing,
Under i.i.d.\ uniform coverage on $[0,1]$, the CDF of the largest spacing is $F({\delta;M})
= \sum_{j=0}^{\lfloor 1/\delta \rfloor}
(-1)^j \binom{M+1}{j}\,(1-j\delta)^M$,
so a one-sided $p$-value for the observed $\delta$ is
\begin{equation}
p_{gap}({\delta;M})=1-F({\delta;M})
=1-\sum_{j=0}^{\lfloor 1/\delta \rfloor}(-1)^j \binom{M+1}{j}\,(1-j\delta)^M .
\label{app:eq:gapPval}
\end{equation}
Small values of $p_{gap}({\delta;M})$ flag clustered sampling along the line (e.g. phase clumping) that can spuriously suggest a steep negative slope without a genuine $\mathcal{SB}2$ signal.

\subsection{Model-selection rules and practical checks}
\label{app:rules}
For each model ($\mathcal{S}1$/$\mathcal{SB}1$/$\mathcal{SB}2$) we compute $S$ (Eq.~\ref{eq:Sdef}), $n_{\rm eff}$ (Eq.~\ref{app:eq:neff}), and BIC (Eq.~\ref{app:eq:bic}) and take the model with the minimal BIC as the raw choice. We then apply simple physics-motivated checks: the Wilson slope must be significantly negative, velocity dispersions should imply sensible semi-amplitudes, and the coverage diagnostic should not be pathological (Eq.~\ref{app:eq:gapPval}). These rules only override the BIC decision in clear-cut edge cases. The semi-amplitudes are estimated as:
\begin{equation}
\widehat{K}_1 = \sqrt{2}\, \hat{\sigma}\{\rm{v}_{1,m};m=1...M\} \, , \qquad
\widehat{K}_2 = \sqrt{2}\, \hat{\sigma}\{\rm{v}_{2,m};m=1...M\} \, ,
\label{app:eq:Kproxies}
\end{equation}
where ${\gamma}$ is the Wilson-fit systemic velocity and $\hat{\sigma}\{\cdot\}$ is the standard deviation operator. Let \(K_{\mathrm{reject}}\) and \(K_{\mathrm{accept}}\) denote the rejection/acceptance amplitude thresholds for the \(\mathcal{SB}1\)/\(\mathcal{SB}2\) classifiers. 
Similarly, \(\left(q/\sigma_q\right)_{\mathrm{reject}}\) and \(\left(q/\sigma_q\right)_{\mathrm{accept}}\) are the rejection/acceptance thresholds for the Wilson-slope significance (used for \(\mathcal{SB}2\)). 
Finally, let \(\varepsilon\) be the minimum \(p_{\mathrm{gap}}\) acceptance threshold for \(\mathcal{SB}2\) (Eq.~\ref{app:eq:gapPval}).
For the LAMOST MRS survey, we adopt the following pragmatic thresholds, tuned to yield robust classification:

\begin{equation}
\begin{aligned}
K_{\mathrm{reject}} &= 5, \quad K_{\mathrm{accept}} = 5,\\
\left(\frac{q}{\sigma_q}\right)_{\mathrm{reject}} &= 5,\quad \left(\frac{q}{\sigma_q}\right)_{\mathrm{accept}} = 12,\quad \varepsilon_{gap} = e^{-25}.
\end{aligned}
\label{app:eq:thr}
\end{equation}

Using these thresholds, the override rules for promoting/demoting the BIC-selected model are:

\begin{enumerate}[leftmargin=*,itemsep=0.6ex]

\item \textit{Promote to \(\mathcal{SB}2\):} If the raw BIC choice is not \(\mathcal{SB}2\), but the Wilson-fit significance and both RV dispersions are high enough, and the phase-coverage diagnostic is well behaved, then promote to \(\mathcal{SB}2\):
\begin{equation}
\begin{aligned}
\mathcal{SB}2 \;\Leftarrow\;&\; \big(S_{\mathrm{BIC}}\neq \mathcal{SB}2\big)\\
&\land\; \Big(\tfrac{q}{\sigma_q}\ge \big(\tfrac{q}{\sigma_q}\big)_{\mathrm{accept}}\Big)
\;\land\; \big(p_{\mathrm{gap}}(\delta;M)>\varepsilon_{gap}\big)
\\[-0.2ex]
&\land\; \big(\widehat K_1\ge K_{\mathrm{accept}}\big)
\;\land\; \big(\widehat K_2\ge K_{\mathrm{accept}}\big).
\end{aligned}
\label{app:eq:override-sb2}
\end{equation}

\item \textit{\(\mathcal{SB}1\) overrides:}
\begin{enumerate}[label=(\alph*),itemsep=0.4ex]

\item \textit{Demote \(\mathcal{SB}2\) to \(\mathcal{SB}1\):} If the raw BIC choice is \(\mathcal{SB}2\) but the Wilson fit is insignificant, demote to \(\mathcal{SB}1\):
\[
\mathcal{SB}1 \;\Leftarrow\;
\big(S_{\mathrm{BIC}}=\mathcal{SB}2\big)
\;\land\;
\Big(\tfrac{q}{\sigma_q}\le \big(\tfrac{q}{\sigma_q}\big)_{\mathrm{reject}}\Big).
\label{app:eq:override-sb1a}
\]

\item \textit{Promote \(\mathcal{S}1\) to \(\mathcal{SB}1\):} If the raw BIC choice is \(\mathcal{S}1\) but the primary RV dispersion is high enough, promote to \(\mathcal{SB}1\):
\[
\mathcal{SB}1 \;\Leftarrow\;
\big(S_{\mathrm{BIC}}=\mathcal{S}1\big)
\;\land\;
\big(\widehat K_1\ge K_{\mathrm{accept}}\big).
\label{app:eq:override-sb1b}
\]
\end{enumerate}

\item \textit{Demote to \(\mathcal{S}1\):} If the raw BIC choice is \(\mathcal{SB}1\) but the primary RV dispersion is too low, demote to \(\mathcal{S}1\):
\[
\mathcal{S}1 \;\Leftarrow\;
\big(S_{\mathrm{BIC}}=\mathcal{SB}1\big)
\;\land\;
\big(\widehat K_1< K_{\mathrm{reject}}\big).
\label{app:eq:override-s1}
\]

\end{enumerate}

\section{Workflow and Software Implementation}
\label{app:workflow}
\setcounter{equation}{0}
\renewcommand{\theequation}{C\arabic{equation}}

This section describes the high-level search workflow with concrete implementation details. It explains how data and templates are prepared, how trials are generated and polished, how per-epoch velocities and $\alpha$ are refined, and how outputs and uncertainties are produced.

\subsection{Inputs and preprocessing}
\label{sec:impl}
The routine \texttt{multiCorr} expects:
(i) a logarithmic wavelength grid $\{\lambda_n\}_{n=1}^{N}$ (constant Doppler step);
(ii) an $M{\times}N$ matrix of continuum–normalized observed spectra;
(iii) per-epoch SNR values;
(iv) a template grid (PHOENIX by default) and allowed bounds in parameter space;
(v) velocity search windows and a coarse step $dV$ (km\,s$^{-1}$\,pix$^{-1}$);
(vi) the instrumental resolving power $R$.

\textbf{Pre-processing of the observed spectra.}
Before any correlation, a fixed, reproducible pipeline is applied to each epoch:
\begin{enumerate}[leftmargin=*,itemsep=0.25ex]
\item \textbf{H$\alpha$ window detection.} The minimum of the H$\alpha$ line near $6562.8$\,\AA\ is located, after which a left/right window is grown until a side maximum exceeds a chosen quantile threshold; that window is held out in subsequent steps.
\item \textbf{Continuum normalization.} Outside the protected H$\alpha$ window, the continuum is fit (least-squares) to the spectrum’s \emph{upper envelope} (short-span moving maxima) on a wavelength axis scaled to $[-1,1]$. The polynomial order is iterated from 2 to 8 while the H$\alpha$ mask is excluded; after each fit, the spectrum is divided by the continuum, residual spikes are suppressed with local MAD rules, and flux is clipped only to broad physical bounds. This yields a flattened spectrum in which lines within the protected H$\alpha$ region are preserved and edge/outlier artifacts are controlled. The continuum is enforced at the spectral edges.
\item \textbf{Resampling to the log-$\lambda$ grid.} The cleaned spectrum is interpolated to the common grid, endpoint pathologies are capped, and global bounds are re-applied.
\item \textbf{Epoch vetting (diagnostic only).} For each epoch, the mean peak height of its 1D-CCF against all other epochs is computed to surface spectra that are too dissimilar for a given target; epochs whose mean-peak statistic is an outlier within the system are flagged for inspection (no automatic rejection).
\end{enumerate}

\textbf{Template preparation}
Templates are drawn from PHOENIX \citep{husser2013}, trimmed to the working band, interpolated, convolved with the instrumental LSF at the target $R$, binned to the survey pixels, and re-normalized. Rotational broadening is applied during the search with the standard kernel $\mathcal{G}(v; \, V\sin{i}, \epsilon)$ (Eq.~\ref{app:eq:rotkernel}), with a fixed limb-darkening coefficient $\epsilon$.

\subsection{Search workflow}
\begin{enumerate}[leftmargin=*,itemsep=0.25ex,topsep=0.4\baselineskip]
\item \textit{Prepare data and templates.} Rectify, mean-subtract, and put all spectra on the common log-$\lambda$ grid. Build templates at the survey resolution and apply rotational broadening.
\item \textit{Initial template guesses.} Generate a wide set of trial parameters that cover the allowed ranges evenly (we use a uniform space-filling draw so the trials are not clustered).
\item \textit{Per-epoch velocities.} For each trial:
  \begin{itemize}[leftmargin=1.2em,itemsep=0.15ex]
  \item $\mathcal{S}1/\mathcal{SB}1$: 1D-CCF $\rightarrow$ $\rm{v}_{1,m}$.
  \item $\mathcal{SB}2$: TODCOR $\rightarrow$ $(\rm{v}_{1,m},\rm{v}_{2,m})$; scan $\alpha$ on a coarse grid.
  \end{itemize}
  Refine peak locations to sub-pixel accuracy (App.~\ref{app:refine}).
\item \textit{Score and keep the best.} Compute $S$ (Section~\ref{sec:multiepoch}, Eq.~\ref{eq:Sdef}) and retain the top-scoring trials.
\item \textit{Local polishing.} Starting from each retained trial, adjust parameters in small steps and re-optimize per-epoch velocities (and $\alpha$ for $\mathcal{SB}2$). When components are similar, also test swapping the primary/secondary assignment; keep only improvements in $S$.
\item \textit{Choose the model.} Compute AIC/BIC with $n_{\rm eff}$ (Eqs.~\ref{app:eq:neff}–\ref{app:eq:bic}) and select $\mathcal{S}1/\mathcal{SB}1/\mathcal{SB}2$. 
\item \textit{Override.} Apply the rules in App.~\ref{app:rules} to select the final model.

\end{enumerate}

\subsection{Model-specific details}
\textbf{$\mathcal{SB}2$ model (double-lined; 8-D template parameters).}
The parameter vector is: \[\theta_{\rm {\scriptstyle{SB2}}}=(T_1,T_2,\log g_1,\log g_2,Z,V\sin{i}_1,V\sin{i}_2,\alpha) \, ,\]
where $\alpha\!=\!F_2/F_1$ is the monochromatic flux ratio.
\begin{enumerate}[leftmargin=*,itemsep=0.25ex]
\item \textbf{Initial trial set in the (7+1)-D box.} Generate a space-filling design via Latin–hypercube sampling (\citealt{McKay1979}): draw \(N_{7D}\) points for the seven non-\(\alpha\) parameters over their allowed ranges, and for each point evaluate \(N_{\alpha}\) uniformly spaced \(\alpha\in[0,1]\) values (defaults: \(N_{7D}=2000\), \(N_{\alpha}=21\)). \label{app:errors:sampling}
\item \textbf{Per-epoch velocities and trial flux ratio.} For each trial point build $(g_1,g_2)$ (including rotational broadening). For all trial $\alpha$, locate the TODCOR maximum to obtain $\{(\hat \rm{v}_{1,m},\hat \rm{v}_{2,m});\, m=1,\ldots,M\}$ and the peak value per epoch, using a coarse lag step (default: $dV\!\sim\!4$\,km\,s$^{-1}$\,pix$^{-1}$). Keep the single $\alpha$ that maximizes the weighted multi-epoch peak strength.
\item \textbf{Local refinement of top trial points.} We refine each leading candidate by (i) a short Levenberg–Marquardt least-squares step using the per-epoch residual vector (\citealp{Marquardt1963}; \citealp{More1978}), followed by (ii) a scalar refinement of the global objective \(1-S^2\) via the Nelder–Mead simplex method (\citealp{NelderMead1965}; \citealp{Lagarias1998}).

\item \textbf{Component-swap tests.} Repeat the local polishing after swapping $(\hat \rm{v}_{1,m},\hat \rm{v}_{2,m})$, and again after swapping the entire $(1\!\leftrightarrow\!2)$ template assignment. Keep a swap only if the multi-epoch score improves.
\item \textbf{Final per-epoch refinement.} With the best \emph{parameters} fixed, jointly re-optimize all per-epoch velocities and $\alpha$ in narrow windows. By convention, if $\alpha\!>\!1$ the components are swapped so that $\alpha\!\le\!1$.
\item \textbf{Wilson fit and diagnostics.} From the refined $(\rm{v}_{1,m},\rm{v}_{2,m})$ fit $\rm{v}_2=a\,\rm{v}_1+b$ ~\citep{York2004} to obtain $(q,\gamma)$ via Eq.~\eqref{app:eq:wgamma}. Also compute the largest-gap $p$-value along the Wilson line (Eq.~\ref{app:eq:gapPval}) to guard against phase clustering.
\end{enumerate}

\textbf{$\mathcal{SB}1$ model (single-lined; 4-D template parameters).}
The parameter vector is $\theta_{\rm SB1}=(T,\log g,Z,V\sin{i})$.
\begin{enumerate}[leftmargin=*,itemsep=0.25ex]
\item \textbf{Initial trial set over the 4-D box.} As above, draw trial points that cover the allowed range, with the same density as in the 7-D grid (default: $2000^{4/7}$ points)
\item \textbf{Per-epoch velocities.} For each trial point build $g_1$ and locate the 1D-CCF maximum per epoch to obtain $\rm{v}_{1,m}$.
\item \textbf{Local polishing and final refinement.} Refine the top trial points locally; then, with the \emph{template parameters} fixed, jointly re-optimize all $\rm{v}_{1,m}$ in narrow windows. Set $\hat{K_1} = \sqrt{2}\,\hat{\sigma}\{\rm{v}_{1,m};m=1...M\}$ as a quick amplitude estimate.
\end{enumerate}

\textbf{$\mathcal{S}1$ model (constant velocity).}
Solve for a single shared velocity across epochs together with a single template. The shared-velocity uncertainty pools per-epoch curvature-based errors by inverse variance.

\subsection{Outputs}
For each model, \emph{MESS} returns:
\begin{itemize}[leftmargin=*,itemsep=0.25ex]
\item best-fit \emph{template parameters} (and $\alpha$ for $\mathcal{SB}2$) with errors and a covariance matrix;
\item per-epoch RVs with uncertainties and peak-correlation values;
\item the global score $S$ and its per-epoch contributions;
\item residual spectra (for diagnostics);
\item AIC/BIC and the effective $n_{\rm eff}$ used;
\item for $\mathcal{SB}2$, the Wilson $(q,\gamma)$ with uncertainties and the largest-gap probability.
\end{itemize}
A \texttt{Summary} block records the selected model, BIC/AIC for all three models, and the diagnostics that triggered any rule-based adjustment that superseded the BIC-only choice.

\section{Synthetic-Data Protocol for Validation}
\label{app:simproto}
\addcontentsline{toc}{section}{Synthetic-data protocol for validation}
\setcounter{equation}{0}
\renewcommand{\theequation}{D\arabic{equation}}

We outline the procedure for generating mock LAMOST-like multi-epoch datasets with known ground truth, which we then analyze end-to-end to benchmark accuracy, precision, and model-selection performance.

\textbf{Parameter draws and scheduling.}
For each realization:
\begin{enumerate}[leftmargin=*,itemsep=0.25ex]
  \item \textit{Model and schedule.} Choose the model uniformly from $\mathcal{S}1$/$\mathcal{SB}1$/$\mathcal{SB}2$, and select observation times $\{t_m\}$ from a survey-like cadence.
  \item \textit{Primary star.} Draw $(T_1,Z)$ and map to $(M_1,R_1)$ via an empirical $T_{\rm eff}$–$Z$ relation; Then obtain  $ \log g_1$ from $M_1$ and $R_1$.
  To relate $(T_{\rm eff},M,R)$, we use the \emph{Modern Mean Dwarf Stellar Color and Effective Temperature Sequence} (ver.~2022.04.16);\footnote{E.~E.~Mamajek, \emph{A Modern Mean Dwarf Stellar Color and Effective Temperature Sequence}, online compilation, version 2022.04.16; \nolinkurl{http://www.pas.rochester.edu/~emamajek/EEM_dwarf_UBVIJHK_colors_Teff.txt}.} see \citet{PecautMamajek2013} and \citet{PecautMamajekBubar2012}.
\item \textit{Secondary and ratios.} For $\mathcal{SB}2$, draw $q$ and $\alpha$ (independently within bounds) and enforce a common $Z$; set $M_2=q\,M_1$.
Let $M_{R_C}$ denote the absolute magnitude in the Cousins-$R$ band (Kron--Cousins system; \citealt{Bessell1990}) from the Mamajek sequence.
When $\alpha$ is specified, we (i) compute the bandpass magnitude shift $\Delta M \equiv -2.5\log_{10}\alpha$; (ii) take the primary’s tabulated $M_{R_C,1}=M_{R_C}(T_1)$ and set $M_{R_C,2}=M_{R_C,1}+\Delta M$; (iii) invert the tabulated relation $M_{R_C}\!\mapsto\!T_{\rm eff}$ to obtain $T_2$; and (iv) infer $R_2$ from $R_1$ using a bandpass-averaged Stefan–Boltzmann approximation.
\end{enumerate}

\textbf{Synthetic LAMOST spectra and processing.}
We generate mock LAMOST MRS red-arm spectra by (i) sampling stellar and orbital parameters and (ii) constructing primary—and, when applicable, secondary—templates from an interpolated, continuum-normalized synthetic grid matched to a LAMOST-like PSF and pixel size, (iii) applying rotational broadening and Doppler shifts on a common log-$\lambda$ grid, and (iv) adding photon-counting (Poisson) noise to reach the desired ${\rm SNR}$. RVs follow Keplerian motion with the eccentric anomaly solved by Newton iterations. For $\mathcal{SB}1$ we set $\alpha{=}0$ (secondary suppressed); for $\mathcal{S}1$ we fix $K_1{=}K_2{=}0$ and a constant RV. If a draw violates the semi-amplitude window, $P$ is slightly adjusted within limits (log-uniform prior) or the draw is rejected to keep $K_1,K_2$ survey-relevant. The pipeline then processes these synthetic spectra exactly as real data.



\textbf{Inference and metrics.}
We run the full workflow (App.~\ref{app:workflow}), calculate per-epoch RV accuracy, and extract $(q,\gamma)$ with uncertainties from the Wilson fit. We then compute BIC for each model, apply the rule-based overrides (App.~\ref{app:rules}), and compile the $\mathcal{S}1/\mathcal{SB}1/\mathcal{SB}2$ confusion matrix (Table~\ref{tab:confusion_sim}). This matrix serves as a primary validation criterion for the algorithm.

\begin{table}
\caption{Distributions used to draw simulated stellar/orbital parameters (truncated where bounds are given). Symbols: $U[a,b]$ uniform on $[a,b]$; $\log U[a,b]$ uniform in $\log$ between $a$ and $b$; $\mathcal{N}(\mu,\sigma)$ normal.}
\label{tab:sim_dists}
\centering
\setlength{\tabcolsep}{6pt}
\renewcommand{\arraystretch}{1.12}
\begin{tabular}{@{}ll@{}}
\toprule
\textbf{Quantity [unit]} & \textbf{Distribution} \\
\midrule
$T_1$ [K]                               & $U[4000,7000]$ \\
$T_2$ [K]                               & $U[3000,7000]$ \\
$\log g_1,\ \log g_2$ [dex]$^{\dagger}$ & implied by $(M,R)$ \\
$Z$ [dex]                               & $\mathcal{N}(-0.05,0.20)$,\ truncated to $[-0.5,0.5]$ \\
$V\sin{i}_1,\ V\sin{i}_2$ [km\,s$^{-1}$]      & $\log U[5,100]$ \\
$\alpha$ $^{\ddagger}$                  & $U[0.01,0.99]$ \\
$q \equiv M_2/M_1$                      & $U[0.25,1]$ \\
$e$                                     & $U[0,0.5]$ \\
$\omega$ [rad]                          & $U[0,2\pi]$ \\
$i$ [rad]                               & isotropic ($\cos i \sim U[-1,1]$) \\
$P$ [d]                                 & $ U[1,100]$ \\
$T_0$ [d]                               & $U[0,P]$ \\
$\gamma$ [km\,s$^{-1}$]                 & $U[-50,50]$ \\
$K_1,\ K_2$ [km\,s$^{-1}$]$^{\S}$       & implied by $(P,q,i,e)$; enforce $[10,50]$ \\
Epochs $\{t_m\}$                        & fixed survey-like schedule$^{\P}$ \\
Noise                                   & Poisson around continuum$^{\ast}$;\ target ${\rm SNR}=50$ \\
\bottomrule
\end{tabular}

\begin{flushleft}
\footnotesize
$^{\dagger}$ From an empirical $T_{\rm eff}\!\to\!(M,R)$ main-sequence mapping.\quad

$^{\ddagger}$ Set to $0$ for $\mathcal{SB}1$/$\mathcal{S}1$.\quad

$^{\S}$ Rejecting draws out of range (or adjust $P$ within limits).\quad

$^{\P}$ Taken from a real LAMOST DR10 survey cadence, for systems with at least ten observation epochs.\quad 

$^{\ast}$ Photon noise with variance $\propto$ flux; see Eq.~(\ref{eq:weights}) for how ${\rm SNR}_m$ and ${\rm Var}(f_m)$ enter the weights.
\end{flushleft}
\end{table}

\endgroup

\end{document}